\RequirePackage{fix-cm}

\documentclass[smallextended]{svjour3}

\smartqed  

\usepackage[square,comma,numbers,sort&compress,sectionbib]{natbib}
\usepackage{times}
\usepackage{graphicx}
\usepackage{amssymb}
\usepackage{tikz}
\usetikzlibrary{decorations}
\usetikzlibrary{decorations.pathmorphing}
\usetikzlibrary{arrows,shapes,positioning,shadows,trees}
\usepackage{epstopdf}
\usetikzlibrary{snakes}
\usepackage{graphicx}
\usepackage{epsfig}
\usepackage{amsmath}
\usepackage{booktabs}
\usepackage{multirow}
\usepackage{tabularx}
\usepackage{tabulary}
\usepackage{booktabs}
\usepackage{subcaption}
\usepackage{acronym}
\usepackage{url}

\begin{document}

\title{A Survey on Neural Models for Web Search}
\title{A Survey on Query-Text Matching in Web Search Using Neural Models}
\title{A Survey on Distributed Representations for Semantic Matching in Web Search}
\title{Getting Started with Neural Models for Semantic Matching in Web Search}



\author{Kezban Dilek Onal \and Ismail Sengor Altingovde \and Pinar Karagoz \and Maarten de Rijke}


\institute{
K.D. Onal \and I.S. Altingovde \and P. Karagoz \at Middle East Technical University \\ \email{dilek@ceng.metu.edu.tr, altingovde@ceng.metu.edu.tr, karagoz@ceng.metu.edu.tr}
\and
K.D. Onal \and M. de Rijke \at University of Amsterdam \\ \email{k.d.onal@uva.nl, derijke@uva.nl}
}

\date{Received: date / Accepted: date}

\maketitle

\begin{abstract}

The vocabulary mismatch problem is a long-standing problem in information retrieval. Semantic matching holds the promise of solving the problem. Recent advances in language technology have given rise to unsupervised neural models for learning representations of words as well as bigger textual units. Such representations enable powerful semantic matching methods. This survey is meant as an introduction to the use of neural models for semantic matching. To remain focused we limit ourselves to web search. We detail the required background and terminology, a taxonomy grouping the rapidly growing body of work in the area, and then survey work on neural models for semantic matching in the context of three tasks: query suggestion, ad retrieval, and document retrieval. We include a section on resources and best practices that we believe will help readers who are new to the area. We conclude with an assessment of the state-of-the-art and suggestions for future work.

\end{abstract}


\section{Introduction}
\label{sec:intro}

In web search, the vocabulary mismatch problem~\cite{furnas-vocabulary-mismatch-1987} necessitates effective semantic similarity functions for textual units of different types. According to \citet{2013_Li_SemanticMatchinginSearch}, semantic matching is concerned with computing the relevance of a document for a query based on representations enriched by linguistic analysis that is meant to capture the semantics of the query and document. In web search, semantic matching is required not only for matching queries to documents but also for matching queries to other textual units. For instance, query suggestions require semantic matching of queries to queries. And retrieving relevant ads for a given query would benefit highly from effective semantic matching. 

Recent advances in language understanding have given rise to neural network models for unsupervised learning of distributed representations of words~\citep{bengio-neural-2003,2013_Mikolov_EfficientEstimationofWordRepresentationsinVectorSpace, 2014_Pennington_Glove:GlobalVectorsforWordRepresentation, 2014_Baroni_Dontcountpredict} and larger textual units~\cite{2016_Hill_LearningDistributedRepresentationsofSentencesfromUnlabelledData, 2014_Le_DistributedRepresentationsofSentencesandDocuments}. A \emph{distributed representation} for a textual unit is a dense real-valued vector that somehow encodes the semantics of the textual unit~\cite{1986_McClelland_Paralleldistributedprocessing}. Distributed representations hold the promise of aiding semantic matching: by mapping words and other textual units to their representations, semantic matches can be computed in the representation space~\citep{2013_Li_SemanticMatchinginSearch}. Indeed, recent improvements in obtaining distributed representations using neural models have quickly been used for semantic matching in web search. 

The problem of mapping words to a representation that can capture their meanings is referred as \emph{distributional semantics} and has been studied for a very long time; see~\cite{2010_Turney_FromFrequencytoMeaning:VectorSpaceModelsofSemantics} for an overview. \emph{Neural language models}, which may be viewed as a particular flavor of distributional semantic models, so-called \emph{context-predicting distributional semantic models}, have been shown to outperform so-called context-counting models such as \ac{HAL}~\cite{hal}, \ac{LSA}~\cite{lsa}, on word analogy and semantic relatedness tasks~\citep{2014_Baroni_Dontcountpredict}. Moreover, \citet{2015_Levy_Improving_Distributional_Similarity_With_Lessons_From_Word_Embeddings} improve context-counting models by adopting lessons from context-predicting models. \citet{bengio-neural-2003} seem to have been the first ones to propose a neural language model; they introduce the idea of simultaneously learning a language model that predicts a word given its context and its representation, a so-called word embedding. This idea has been since been adopted by many follow-up studies. The most well-known and most widely used context-predicting models, Word2Vec~\citep{2013_Mikolov_EfficientEstimationofWordRepresentationsinVectorSpace} and GloVe \cite{2014_Pennington_Glove:GlobalVectorsforWordRepresentation}, have been used extensively in recent work on web search. The success of neural word embeddings has also given rise to work on computing context-predicting representations of larger textual units, including paragraphs and documents~\citep{2016_Hill_LearningDistributedRepresentationsofSentencesfromUnlabelledData}. 

In recent years, neural network-based models have given rise to significant performance improvements in computer vision, speech processing and machine translation~\cite{deng-deep-2013}. Although the well-known training algorithm back-propagation dates back to 1980s, training deep neural networks was not possible due to optimization issues such as vanishing gradient and local optima. Training deep neural networks to learn hierarchies of representations became possible owing both to theoretical contributions of new machine learning algorithms and parallelization of training on GPUs~\cite{deng-deep-2013}.

Inspired by the success of neural models in other areas of computer science and artificial intelligence, there has been a tremendous growth in interest in using neural models for semantic matching in information retrieval. Despite the growing literature and a set of emerging patterns, to the best of our knowledge, there is no proper introduction to this emerging area. \citet{deng-deep-2013} devote a chapter to applications of deep learning in information retrieval. However, this chapter only covers very early work that is mostly focused on document representations for retrieval. The recent tutorial on deep learning for information retrieval by~\citet{2016_Li_Sigir_Tutorial_DL_for_IR} sketches a range of potential applications of deep learning to \ac{IR} problems with a broader scope.

The purpose of this survey is to provide a systematic introduction to the emerging area of neural models for semantic matching. While the area is still growing rapidly, by now a solid body of shared methods and architectures has been established, which we believe justifies a systematic introduction. We review the literature in the area and consider both word representations and representations of larger textual units, such as sentences and paragraphs. We survey the use of neural language models and word embeddings, we detail applications of so-called neural semantic compositionality models that determine semantic representations of larger text units from smaller ones, and we present novel neural semantic compositionality models designed specifically for web search tasks. In addition to surveying relevant literature, we provide the reader with a foundation of on neural language models and with pointers to relevant resources that those who are new to the area should appreciate.

To remain focused we limit the scope of this survey to the use of neural models for building distributed representations of textual units in web search. Therefore, studies that adopt neural models for other information retrieval tasks, such as semantic expertise retrieval~\cite{vangysel-unsupervised-2016}, product search~\cite{vangysel-learning-2016}, click and behavior models~\cite{borisov-neural-2016,borisov-context-aware-2016}, text similarity~\cite{2015_Kenter_ShortTextSimilaritywithWordEmbeddings,kenter-ad-hoc-2015}, subtopic mining~\cite{2014_Song_CNUSysteminNTCIR-11IMineTask}, are out of scope. Specifically, we focus on three major web search tasks, namely query suggestion, ad retrieval and document retrieval. As we will see, each task requires semantic matching of different types of textual units. 

Section~\ref{sec:adhoc}, \ref{sec:query}, and \ref{sec:ads} are devoted to studies on document retrieval, query suggestion and ad retrieval in web search, respectively. Prior to this, we introduce relevant background information on neural language models in Section~\ref{sec:background}, survey available resources in Section~\ref{sec:resources}, and detail the dimensions that we use for categorizing the publications we review in Section~\ref{sec:taxonomy}. 


\section{Background and terminology}
\label{sec:background}

In this section we briefly review a number of relevant key concepts underlying the use of neural models for semantic matching, viz.\ distributional semantics, semantic compositionality, neural language models, training procedures for neural language models, Word2Vec and GloVe, and paragraph vectors.

We assume that the reader has a basic understanding of neural networks (including RNNs, CNNs, LSTMs), back propagation, forward propagation, and gradient descent as can be found in, e.g., Part~II of~\cite{Goodfellow-et-al-2016-Book}.

\subsection{Distributional semantics}

A \acfi{DSM}\acused{DSM} is a model that relies on the \emph{distributional hypothesis}~\citep{Distributional-Hypothesis}, according to which \emph{words that occur in the same contexts tend to have similar meanings}, for associating words with vectors that can capture their                                                                                                                                                                                                                                                                                                                                                                                                                                                                                                                                                                                                                                                                                                                                                                                                                                                                                                                                                                                                                                                                                                                                                                                                                                                                                                                                                                                                                                                                                                                                                                                                                                                                                                                                                                                                                                                                                                                                                                                                                                                                                                                                                                                                                                                                                                                                                                                                                                                                                                                                                                                                                                                                                                                                                                                                                                                                                                                                                                                                                                                                                                                                                                                                                                                                                                                                                                                                                                                                                                                                                                                                                                                                                                                                                                                                    meaning. Statistics on observed contexts of words in a corpus is quantified to derive word vectors. The most common choice of context is the set of words that co-occur in a context window. 

\citet{2014_Baroni_Dontcountpredict} classify existing \acp{DSM} into two categories:  \emph{context-counting} and \emph{context-predicting}. The context-counting category includes earlier DSMs such as \acf{HAL}~\cite{hal}, \acf{LSA}~\cite{lsa}. In these models, low-dimensional word vectors are obtained via factorisation of a high-dimensional sparse co-occurrence matrix. The \emph{context-predicting} models are neural language models in which word vectors are modelled as additional parameters of a neural network that predicts co-occurrence likelihood of context-word pairs. Neural language models comprise an embedding layer that maps a word to its distributed representation. A \emph{distributed representation} of a symbol is a vector of features that characterize the meaning of the symbol and are not mutually exclusive \cite{1986_McClelland_Paralleldistributedprocessing}. 

Early neural language models were not aimed at learning representations for words. However, it soon turned out that the embedding layer component, which addresses the curse of dimensionality caused by one-hot vectors \citep{bengio-neural-2003}, yields useful distributed word representations, so-called \emph{word embeddings}. \citet{2008_Collobert_AUnifiedArchitectureforNaturalLanguageProcessing:DeepNeuralNetworkswithMultitaskLearning} are the first ones to show the benefit of word embeddings as features for NLP tasks. Subsequently, word embeddings became widespread after introduction of the shallow models Skip-Gram and CBOW in the \emph{Word2Vec} framework by  \citet{2013_Mikolov_EfficientEstimationofWordRepresentationsinVectorSpace, 2013_Mikolov_DistributedRepresentationsofWordsandPhrasesandtheirCompositionality}; see Section~\ref{sec:word2vec}.

\citet{2014_Baroni_Dontcountpredict} report that context-predicting models outperform context-counting models on several tasks including question sets, semantic relatedness, synonym detection, concept categorization and word analogy. In contrast, \citet{2015_Levy_Improving_Distributional_Similarity_With_Lessons_From_Word_Embeddings} point out that the success of the popular context-predicting models Word2Vec and GloVe does not originate from the neural network architecture and the training objective but from the choices of hyper-parameters for contexts. A comprehensive analysis reveals that when these hyper-parameter choices are applied to context-counting models, no consistent advantage of context-predicting models is observed over context-counting models. To sum up, it is possible to obtain word vectors that can encode semantics successfully, owing to the contributions brought by neural language models. 


\subsection{Semantic compositionality}

\emph{Compositional distributional semantics} or \acfi{SC}\acused{SC} is the problem of formalizing how the meaning of larger textual units such as sentences, phrases, paragraphs and documents are built from the meanings of words~\cite{2010_Mitchell_CompositioninDistributionalModelsofSemantics}. Work on \acs{SC} studies are motivated by the \emph{Principle of Compositionality} which states that the meaning of a complex expression is determined by the meanings of its constituent expressions and the rules used to combine them.


A \emph{neural \acs{SC}} model maps the high-dimensional representation of a textual unit into a distributed representation by forward propagation in a neural network. The neural network parameters are learned by training to optimize task-specific objectives. Both the granularity of the target textual unit and the target task play an important role for the choice of neural network type and training objective. An \ac{SC} model that considers the order of words in a sentence and aims to obtain a deep understanding may fail in an application that requires representations that can encode high-level concepts in a large document. A comparison of neural \ac{SC} models of sentences learned from unlabelled data is presented  in \cite{2016_Hill_LearningDistributedRepresentationsofSentencesfromUnlabelledData}. Besides the models reviewed by \citet{2016_Hill_LearningDistributedRepresentationsofSentencesfromUnlabelledData}, there exist sentence-level models that are trained using task-specific labelled data. For instance, a model can be trained to encode the sentiment of a sentence~using a dataset of sentences annotated with sentiment class labels~\cite{2015_Li_When_Are_Tree_Structures_Necessary_for_Deep_Learning_of_Representations}.

To the best of our knowledge, there is no survey on neural \ac{SC} models for distributed representations of long documents, although the representations are useful not only for document retrieval but also for document classification and recommendation. In Section~\ref{sec:tax-learn} we review the subset of neural \ac{SC} models and associated training objectives adopted specifically in web search tasks. 

\subsection{Neural language models}
\label{subsec:nlm}

A \emph{language model} is a function that predicts the acceptability of pieces of text in a language. Acceptability scores are useful for ranking candidates in tasks like machine translation or speech recognition. The probability of a sequence of words  $P(w_1,w_2$, \ldots, $w_n)$ in a language, can be computed by Equation \ref{eq:prob-lm}, in accordance with the chain rule:
\begin{equation}\label{eq:prob-lm}
	P(w_1,w_2,\ldots,w_{t-1},w_t)=P(w_1)P(w_2\mid w_1)\cdots P(w_t\mid w_1,w_2,\ldots,w_{t-1})
\end{equation}
Probabilistic language models, mostly approximate Equation~\ref{eq:prob-lm} by 
$P(w_t\mid w_{t-n}$, \ldots, $w_{t-1})$, considering only a limited context of size $n$, that immediately precedes $w_t$. In neural language models the probability $P(w \mid c)$ of a word $w$ to follow the context $c$ is computed by a neural network. The neural network takes a context $c$ and outputs the conditional probability $P(w \mid c)$ of every word $w$ in the vocabulary $V$ of the language:
\begin{equation}\label{eq:nlm-prob-dist}
     P(w \mid c,\theta) = \frac{\exp(s_\theta(w,c))}{\sum_{w' \in V} \exp(s_\theta(w',c))}.
\end{equation}
Here, $s_\theta(w,c)$ is an unnormalized score for the compatibility of $w$ given the context $c$; $s_\theta(w,c)$ is computed via forward propagation of the context $c$ through a neural network defined with the set of parameters $\theta$. Note that $P(w \mid c)$ is computed by the normalized exponential (softmax) function in Equation \ref{eq:nlm-prob-dist}, over the $s_\theta(w,c)$ scores for the entire vocabulary.

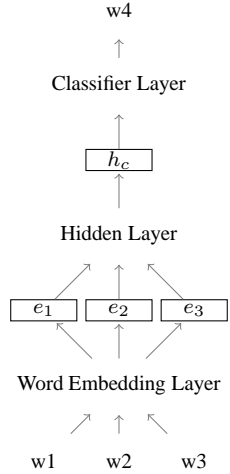
\begin{figure}[t]
\centering

\def\layersep{1cm}
\begin{tikzpicture}[shorten >=1pt,->,draw=black!50, node distance=\layersep,transform shape,rotate=90]  
    \tikzstyle{every pin edge}=[<-,shorten <=1pt]
    \tikzstyle{neuron}=[circle,draw=black,minimum size=17pt,inner sep=0pt]
    \tikzstyle{vector}=[rectangle,draw=black,minimum width=8pt, minimum height=25pt,inner sep=0pt]
    \tikzstyle{layer}=[rectangle,draw=black,minimum width=5pt, minimum height=50pt,inner sep=0pt]
    \tikzstyle{subnet}=[rectangle,draw=black,minimum size=20pt,inner sep=0pt]
    \tikzstyle{justtex}=[rectangle,minimum size=20pt,inner sep=0pt]
    \tikzstyle{input neuron}=[neuron, fill=green!50];
    \tikzstyle{output neuron}=[neuron, fill=red!50];
    \tikzstyle{hidden neuron}=[neuron, fill=blue!50];
    \tikzstyle{annot} = [text width=4em, text centered]
    \tikzset{hoz/.style={rotate=-90}}   
    
   \path node[justtex,hoz,node distance=2cm] (I-1) at (0,3) {w1};
   \path node[justtex,hoz,node distance=2cm] (I-2) at (0, 2) {w2};
   \path node[justtex,hoz,node distance=2cm] (I-3) at (0, 1) {w3};
   \path node[justtex,hoz,node distance=2cm] (E-L) at (1*\layersep, 2) {Word Embedding Layer};
   
   \path node[justtex,hoz,right of=I-5, node distance=2cm] (L-I) at (4*\layersep,4) {$h_c$};
   \path node[justtex,hoz,right of=I-5, node distance=2cm] (L-I) at (2*\layersep,5) {$e_1$};
   \path node[justtex,hoz,right of=I-5, node distance=2cm] (L-I) at (2*\layersep,4) {$e_2$};
   \path node[justtex,hoz,right of=I-5, node distance=2cm] (L-I) at (2*\layersep,3) {$e_3$};
   
   \foreach \source in {1,...,3}
            \path (I-\source) edge (E-L);

    \foreach \name / \y in {1,...,3}
        \path node[vector] (E-\name) at (2*\layersep,\y) {};

   \foreach \source in {1,...,3}
            \path (E-L) edge (E-\source);
        
   \path node[justtex,hoz,node distance=2cm] (H-L) at (3*\layersep, 2) {Hidden Layer};
   
   \path node[vector] (H-V) at (4*\layersep,2) {};
   \path (H-L) edge (H-V);

    \foreach \source in {1,...,3}
            \path (E-\source) edge (H-L);
            
   \path node[justtex,hoz,node distance=2cm] (C-L) at (5*\layersep, 2) {Classifier Layer};
               \path (H-V) edge (C-L);

    \path node[justtex,hoz,node distance=2cm] (O) at (6*\layersep,2) {w4};
    \path (C-L) edge (O);
\end{tikzpicture}
\caption{Architecture of neural language models.}
\label{fig:nlm}
\end{figure}

Parameters of the neural network are learned by training on a text corpus using gradient-descent based optimization algorithms to maximize the likelihood function $L$ in Equation \ref{eq:nlm-loss}, on a corpus: 
\begin{equation}\label{eq:nlm-loss}
     L(\theta) = \sum_{(t,c) \in T} P(t \mid c, \theta)
\end{equation}
The first neural language model published is the \ac{NNLM}~\citep{bengio-neural-2003}. The common architecture shared by neural language models is depicted in Figure \ref{fig:nlm}, with example input context $c=w_1,w_2,w_3$ and the word to predict being $w_4$, extracted from the observed sequence $c=w_1,w_2,w_3,w_4$. Although a probability distribution over the vocabulary $V$ is computed, the word that should have the maximum probability is shown at the output layer, for illustration purposes. 

The neural network takes one-hot vectors $w_1,w_2,w_3$ of the words in the context. The dimensionality of the one-hot vectors is $1 \times |V|$. The \emph{embedding layer} $E$ in Figure~\ref{fig:nlm} is indeed a $|V| \times d$-dimensional matrix whose $i$-th row is the $d$-dimensional word embedding for the $i$-th word in the vocabulary. The embedding vector of the $i$-th word in the vocabulary is obtained by multiplying the one-hot vector of the word with the $E$ matrix or simply extracting the $i_{th}$ row of the embedding matrix. Consequently, high dimensional one-hot vectors of words $w_1,w_2,w_3$ are mapped to their $d$-dimensional embedding vectors $e_1,e_2,e_3$ by the embedding layer. Note that $d$ is usually chosen to be in the range 100--500 whereas $|V|$ can go up to millions. 

The \emph{hidden layer} in Figure~\ref{fig:nlm} takes the embedding vectors $e_1,e_2,e_3$ of the context words and creates a vector $h_c$ for the input context. This layer differs between neural language model architectures. In \ac{NNLM}~\citep{bengio-neural-2003} it is a non-linear neural network layer whereas in the \ac{CBOW} model of Word2Vec~\citep{2013_Mikolov_EfficientEstimationofWordRepresentationsinVectorSpace}, it is vector addition over word embeddings. In the \ac{RNNLM}~\citep{2010_Mikolov_Recurrentneuralnetworkbasedlanguagemodel} the hidden context representation is computed by a recurrent neural network. Besides the hidden layer, the choice of context also differs among models. In \citep{2008_Collobert_AUnifiedArchitectureforNaturalLanguageProcessing:DeepNeuralNetworkswithMultitaskLearning} and in the CBOW model \citep{2013_Mikolov_EfficientEstimationofWordRepresentationsinVectorSpace} context is defined by the words that surround a center word in a symmetric context. In the \ac{NNLM}~\cite{bengio-neural-2003} and \ac{RNNLM}~\citep{2010_Mikolov_Recurrentneuralnetworkbasedlanguagemodel} models, the context is defined by words that precede the target word. The Skip-Gram \cite{2013_Mikolov_EfficientEstimationofWordRepresentationsinVectorSpace} takes a single word as input and predict words from a dynamically sized symmetric context window around the input word.

The \emph{classifier layer} in Figure~\ref{fig:nlm}, which is composed of a weights matrix $C$ of dimension $d \times |V|$ and a bias vector of dimension $|V|$, is used to compute $s_\theta(w,c)$ using Equation~\ref{eq:sscore}:
\begin{equation}\label{eq:sscore}
     s_\theta(w,c) = h_c C + b.
\end{equation}
To sum up, the neural network architecture for a \ac{NLM} is defined by $|V|$, $d$, the context type, the context size $|c|$ and the function in the hidden layer. The parameter set $\theta$ to be optimized includes the embedding matrix $E$, parameters from the hidden layer, the weights matrix $C$ and the bias vector $b$ of the classifier layer. The embedding layer $E$ is treated as an ordinary layer of the network, its weights are initialized randomly and updated with back-propagation during training of the neural network.

 
\subsection{Efficient training of NLMs}
\label{subsec:eff-nlm}

As mentioned previously in our discussion of Equation \ref{eq:prob-lm}, the output of a \ac{NLM} is a normalized probability distribution over the entire vocabulary. Therefore, for each training sample (context pair $(t,c)$), it is necessary to compute the softmax function in Equation~\ref{eq:nlm-prob-dist} and consider the whole vocabulary for computing the gradients of the likelihood function in back-propagation. This makes the training procedure computationally expensive and prevents the scalability of the models to very large corpora. 

Several remedies for efficiently training \acp{NLM} have been introduced. \citet{Chen_Strategies_For_Training_Large_Vocabulary_NLMs_2015} present a comparison of training methods for the \ac{NNLM}. Hierarchical Softmax~\citep{2005_Morin_HierarchicalProbabilisticNeuralNetworkLanguageModel} and differentiated softmax~\cite{Chen_Strategies_For_Training_Large_Vocabulary_NLMs_2015} propose adapted softmax layer architectures for efficient computation of the softmax function. Another solution, adopted by methods like \ac{IS}~\citep{Bengio_Importance_Sampling_for_NLM_2003} and \ac{NCE}~\citep{2012_Mnih_Fast_Simle_Training_NLM}, is to avoid the normalization by using modified loss functions to approximate the softmax. \citet{2008_Collobert_AUnifiedArchitectureforNaturalLanguageProcessing:DeepNeuralNetworkswithMultitaskLearning} propose the cost function in Equation~\ref{eq:collobert-cost}, which does not require normalization over the vocabulary. The \ac{NLM} is trained to compute higher $s_\theta$ scores for observed context-word pairs $(c,t)$ compared to the negative samples constructed by replacing $t$ with any other word $w$ in $V$.  The context is defined as the words in a symmetric window around the center word $t$.
\begin{equation} \label{eq:collobert-cost}
 \sum_{(t,c) \in T} \sum_{w \in V} \max(0, 1 - s_\theta(t,c) + s_\theta(w,c))
\end{equation}
\citet{2012_Mnih_Fast_Simle_Training_NLM} apply \ac{NCE}~\cite{2012_Gutman_Noise_Contrastive_Estimation} to \ac{NLM} training. By using \ac{NCE}, the probability density estimation problem is converted to a binary classification problem. A two-class training data set is created from the training corpus by treating the observed context-word pairs $(t,c)$ as positive samples and noisy pairs $(t',c)$ constructed replacing $t$ with a word $t'$ sampled from the noise distribution $q$. The \ac{NCE} cost function defined for $k$ negative samples per observed sample is given in Equation~\ref{eq:nce-cost}, where conditional probabilities of classes are computed as in Equation~\ref{eq:nce-cond-prob}:
\begin{eqnarray} 
L_{\mathit{NCE}_k} &=& \sum_{(t,c) \in T} \left( \log p(l=1\mid t,c) + \sum_{i=1,t \sim q}{k} \log p(l=0\mid t',c) \right) 
\label{eq:nce-cost} \\
p(l=0\mid w,c) &=& \frac{k \times q(w)}{ s_\theta(w,c) + k \times q(w)}
\label{eq:nce-cond-prob} \\
p(l=1\mid w,c) &=& \frac{s_\theta(w,c)}{ s_\theta(w,c) + k \times q(w)}.
\end{eqnarray}

\subsection{Word2Vec and GloVe}
\label{sec:word2vec}

\begin{figure}[t]
  \centering
  \begin{minipage}[b]{0.45\textwidth}
    \centering
    \begin{tikzpicture}[shorten >=1pt,->,draw=black!50, node distance=\layersep,transform shape,rotate=90]  
    \tikzstyle{every pin edge}=[<-,shorten <=1pt]
    \tikzstyle{neuron}=[circle,draw=black,minimum size=17pt,inner sep=0pt]
    \tikzstyle{vector}=[rectangle,draw=black,minimum width=5pt, minimum height=20pt,inner sep=0pt]
    \tikzstyle{layer}=[rectangle,draw=black,minimum width=5pt, minimum height=50pt,inner sep=0pt]
    \tikzstyle{subnet}=[rectangle,draw=black,minimum size=20pt,inner sep=0pt]
    \tikzstyle{justtex}=[rectangle,minimum size=20pt,inner sep=0pt]
    \tikzstyle{input neuron}=[neuron, fill=green!50];
    \tikzstyle{output neuron}=[neuron, fill=red!50];
    \tikzstyle{hidden neuron}=[neuron, fill=blue!50];
    \tikzstyle{annot} = [text width=4em, text centered]
    \tikzset{hoz/.style={rotate=-90}}   
    
   \path node[justtex,hoz,node distance=2cm] (I-1) at (0,3.5) {$w_1$};
   \path node[justtex,hoz,node distance=2cm] (I-2) at (0,2.5) {$w_2$};
   \path node[justtex,hoz,node distance=2cm] (I-3) at (0,1.5) {$w_4$};
   \path node[justtex,hoz,node distance=2cm] (I-4) at (0,0.5) {$w_5$};
   \path node[justtex,hoz,node distance=2cm] (O) at (4, 2) {$w_3$};
   \path node[justtex,hoz,node distance=2cm] (E-L) at (1, 2) {Word Embedding Layer ($E$)};

   \foreach \source in {1,...,4}
            \path (I-\source) edge (E-L);

   \path node[vector] (E-1) at (2,3.5){};
   \path node[vector] (E-2) at (2,2.5){};
   \path node[vector] (E-3) at (2,1.5){};
   \path node[vector] (E-4) at (2,0.5) {};

   \foreach \source in {1,...,4}
            \path (E-L) edge (E-\source);
   
   \path node[vector] (H-V) at (3,2) {};
   \foreach \source in {1,...,4}
            \path (E-\source) edge (H-V);
    \path (H-V) edge (O);
\end{tikzpicture}
    \subcaption{Continuous Bag of Words (CBOW)}
    \label{fig:cbow}
  \end{minipage}
  \hfill
  \begin{minipage}[b]{0.45\textwidth}
    \centering
    \begin{tikzpicture}[shorten >=1pt,->,draw=black!50, node distance=\layersep,transform shape,rotate=90]  
    \tikzstyle{every pin edge}=[<-,shorten <=1pt]
    \tikzstyle{neuron}=[circle,draw=black,minimum size=17pt,inner sep=0pt]
    \tikzstyle{vector}=[rectangle,draw=black,minimum width=5pt, minimum height=20pt,inner sep=0pt]
    \tikzstyle{layer}=[rectangle,draw=black,minimum width=5pt, minimum height=50pt,inner sep=0pt]
    \tikzstyle{subnet}=[rectangle,draw=black,minimum size=20pt,inner sep=0pt]
    \tikzstyle{justtex}=[rectangle,minimum size=20pt,inner sep=0pt]
    \tikzstyle{input neuron}=[neuron, fill=green!50];
    \tikzstyle{output neuron}=[neuron, fill=red!50];
    \tikzstyle{hidden neuron}=[neuron, fill=blue!50];
    \tikzstyle{annot} = [text width=4em, text centered]
    \tikzset{hoz/.style={rotate=-90}}   
   
   \path node[justtex,hoz,node distance=2cm] (O-1) at (3,3.5) {$w_1$};
   \path node[justtex,hoz,node distance=2cm] (O-2) at (3,2.5) {$w_2$};
   \path node[justtex,hoz,node distance=2cm] (O-3) at (3,1.5) {$w_4$};
   \path node[justtex,hoz,node distance=2cm] (O-4) at (3,0.5) {$w_5$};
   \path node[justtex,hoz,node distance=2cm] (D) at (0, 2) {$w3$};
   \path node[justtex,hoz,node distance=2cm] (PEM) at (1, 2) {Word Embedding Layer ($E$)};
   \path node[vector] (E-P) at (2,2) {};
   \path (D) edge (PEM);
   \path (PEM) edge (E-P);
   \foreach \source in {1,...,4}
       \path (E-P) edge (O-\source);
   
\end{tikzpicture}
    \subcaption{Skip-Gram}
    \label{fig:skipgram}
  \end{minipage}
  \caption{Models of the Word2Vec framework.}
  \label{fig:w2vec}
\end{figure}
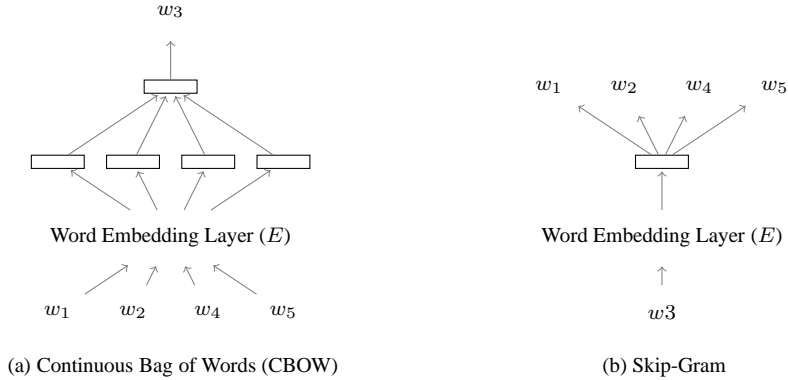

\citet{2013_Mikolov_EfficientEstimationofWordRepresentationsinVectorSpace} introduce the Skip-Gram and CBOW models that follow the NLM architecture with a linear layer for computing a distributed context representation. Figure~\ref{fig:w2vec} illustrate the architecture of Word2Vec models with context windows of size five. The CBOW model is trained to predict the center word of a given context. In the CBOW model, the hidden context representation is computed by the sum of the word embeddings. On the contrary, the Skip-Gram model is trained to predict words that occur in a symmetric context window given the center word. The name Skip-Gram is used since the size of the symmetric context window is sampled randomly from the range $[0,c]$ for each word. Skip-Gram embeddings are shown to outperform embeddings obtained from \acp{NNLM} and \acp{RNNLM} in capturing the semantic and syntactic relationships between the words. 


Efficient training of Skip-Gram and CBOW models is achieved by hiearchical softmax~\cite{2005_Morin_HierarchicalProbabilisticNeuralNetworkLanguageModel} with a Huffman tree~\citet{2013_Mikolov_EfficientEstimationofWordRepresentationsinVectorSpace}. In follow-up work~\cite{2013_Mikolov_DistributedRepresentationsofWordsandPhrasesandtheirCompositionality}, \ac{NEG} is proposed for efficiently training the Skip-Gram model. \ac{NEG} is a variant of \ac{NCE} in which conditional probabilities are computed by Equation~\ref{eq:ns-prob}. \ac{NEG} adopts the loss function in Equation~\ref{eq:nce-cost} but the noise distribution $q$ is assumed to be uniform and $k=|V|$ while computing the conditional probabilities. \ac{SGNS} departs from the goal of learning a language model and is biased towards the quality word embeddings. \ac{SGNS} is never used as a language model, therefore it should be considered to be a \ac{DSM} rather than a language model:
\begin{eqnarray} 
\label{eq:ns-prob}
p(l=0 \mid w,c) &=& \frac{1}{ s_\theta(w,c) + 1}\\
 p(l=1\mid w,c) &=& \frac{s_\theta(w,c)}{ s_\theta(w,c) +1}
\end{eqnarray}
Subsampling frequent words is another extension introduced in~\cite{2013_Mikolov_DistributedRepresentationsofWordsandPhrasesandtheirCompositionality} for speeding up training and increasing the quality of embeddings. Each word $w_i$ in the corpus is discarded with probability $p(w_i)$, given in Equation~\ref{eq:sg-subsample}: 
\begin{equation}\label{eq:sg-subsample}
p(w_i) = 1 - \sqrt{\frac{t}{f(w_i)}}.
\end{equation}
Global Vectors (GloVe)~\cite{2014_Pennington_Glove:GlobalVectorsforWordRepresentation}  combines global context and local context in the training objective for learning word embeddings. In contrast to \acp{NLM}, where embeddings are optimized to maximize the likelihood of local contexts, GloVe embeddings are trained to fit the co-occurrence ratio matrix. 

\citet{2015_Levy_Improving_Distributional_Similarity_With_Lessons_From_Word_Embeddings} discuss that diluting frequent words before training enlarges the context window size in practice. Experiments show that the hyper-parameters about context-windows like dynamic size and subsampling frequent words have a notable impact on the performance of \ac{SGNS} and GloVe~\cite{2015_Levy_Improving_Distributional_Similarity_With_Lessons_From_Word_Embeddings}. \citeauthor{2015_Levy_Improving_Distributional_Similarity_With_Lessons_From_Word_Embeddings} show that when these choices are applied to traditional DSMs, no consistent advantage of \ac{SGNS} and GloVe is observed. In contrast to the conclusions obtained in~\citep{2014_Baroni_Dontcountpredict}, the success of context-predicting models is attributed to choice of hyper-parameters, which can also be used for context-counting \acp{DSM}, rather than to the neural architecture or the training objective. 

\subsection{Paragraph vector}

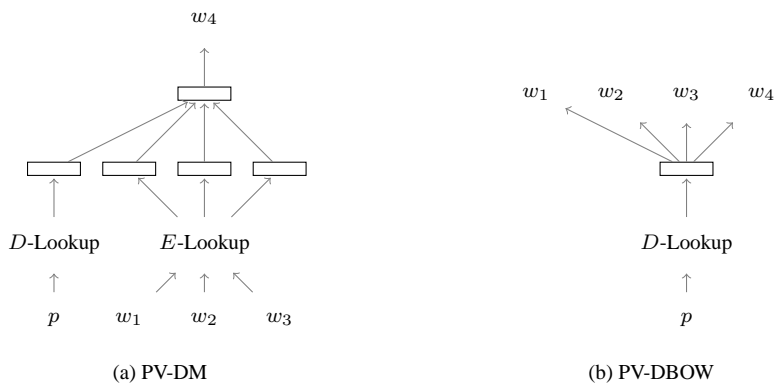
\begin{figure}[t]
  \centering
  \begin{minipage}[b]{0.45\textwidth}
    \centering
    \begin{tikzpicture}[shorten >=1pt,->,draw=black!50, node distance=\layersep,transform shape,rotate=90]  
    \tikzstyle{every pin edge}=[<-,shorten <=1pt]
    \tikzstyle{neuron}=[circle,draw=black,minimum size=17pt,inner sep=0pt]
    \tikzstyle{vector}=[rectangle,draw=black,minimum width=5pt, minimum height=20pt,inner sep=0pt]
    \tikzstyle{layer}=[rectangle,draw=black,minimum width=5pt, minimum height=50pt,inner sep=0pt]
    \tikzstyle{subnet}=[rectangle,draw=black,minimum size=20pt,inner sep=0pt]
    \tikzstyle{justtex}=[rectangle,minimum size=20pt,inner sep=0pt]
    \tikzstyle{input neuron}=[neuron, fill=green!50];
    \tikzstyle{output neuron}=[neuron, fill=red!50];
    \tikzstyle{hidden neuron}=[neuron, fill=blue!50];
    \tikzstyle{annot} = [text width=4em, text centered]
    \tikzset{hoz/.style={rotate=-90}}   
    
   \path node[justtex,hoz,node distance=2cm] (I-1) at (0,3) {$w_1$};
   \path node[justtex,hoz,node distance=2cm] (I-2) at (0, 2) {$w_2$};
   \path node[justtex,hoz,node distance=2cm] (I-3) at (0, 1) {$w_3$};
      \path node[justtex,hoz,node distance=2cm] (O) at (4, 2) {$w_4$};

   \path node[justtex,hoz,node distance=2cm] (WEM) at (1, 2) {$E$-Lookup};
   \path node[justtex,hoz,node distance=2cm] (D) at (0, 4) {$p$};
   \path node[justtex,hoz,node distance=2cm] (PEM) at (1, 4) {$D$-Lookup};
   \path (D) edge (PEM);
   \path node[vector] (E-P) at (2,4) {};
   \path (PEM) edge (E-P);

   \foreach \source in {1,...,3}
            \path (I-\source) edge (E-L);

    \foreach \name / \y in {1,...,3}
        \path node[vector] (E-\name) at (2,\y) {};

   \foreach \source in {1,...,3}
            \path (E-L) edge (E-\source);
   
   \path node[vector] (H-V) at (3,2) {};
   \path (E-P) edge (H-V);
   \foreach \source in {1,...,3}
            \path (E-\source) edge (H-V);
    \path (H-V) edge (O);
\end{tikzpicture}
    \subcaption{PV-DM}
    \label{fig:pv-dm}
  \end{minipage}
  \hfill
  \begin{minipage}[b]{0.45\textwidth}
    \centering
    \begin{tikzpicture}[shorten >=1pt,->,draw=black!50, node distance=\layersep,transform shape,rotate=90]  
    \tikzstyle{every pin edge}=[<-,shorten <=1pt]
    \tikzstyle{neuron}=[circle,draw=black,minimum size=17pt,inner sep=0pt]
    \tikzstyle{vector}=[rectangle,draw=black,minimum width=5pt, minimum height=20pt,inner sep=0pt]
    \tikzstyle{layer}=[rectangle,draw=black,minimum width=5pt, minimum height=50pt,inner sep=0pt]
    \tikzstyle{subnet}=[rectangle,draw=black,minimum size=20pt,inner sep=0pt]
    \tikzstyle{justtex}=[rectangle,minimum size=20pt,inner sep=0pt]
    \tikzstyle{input neuron}=[neuron, fill=green!50];
    \tikzstyle{output neuron}=[neuron, fill=red!50];
    \tikzstyle{hidden neuron}=[neuron, fill=blue!50];
    \tikzstyle{annot} = [text width=4em, text centered]
    \tikzset{hoz/.style={rotate=-90}}   
   
   \path node[justtex,hoz,node distance=2cm] (O-1) at (3,4) {$w_1$};
   \path node[justtex,hoz,node distance=2cm] (O-2) at (3,3) {$w_2$};
   \path node[justtex,hoz,node distance=2cm] (O-3) at (3,2) {$w_3$};
   \path node[justtex,hoz,node distance=2cm] (O-4) at (3,1) {$w_4$};
   
   \path node[justtex,hoz,node distance=2cm] (D) at (0, 2) {$p$};
   \path node[justtex,hoz,node distance=2cm] (PEM) at (1, 2) {$D$-Lookup};
   \path node[vector] (E-P) at (2,2) {};

   \path (D) edge (PEM);

   \path (PEM) edge (E-P);
   \foreach \source in {1,...,4}
       \path (E-P) edge (O-\source);
   
\end{tikzpicture}
    \subcaption{PV-DBOW}
    \label{fig:pv-dbow}
  \end{minipage}
  \caption{Models of the paragraph vector framework.}
  \label{fig:pv}
\end{figure}

The Paragraph Vector~\cite{2014_Le_DistributedRepresentationsofSentencesandDocuments} extends Word2Vec in order to learn representations for so-called \emph{paragraph}, textual units of any length. Similar to Word2Vec, it is composed of two separate models, namely \ac{PV-DM} and \ac{PV-DBOW}. The architectures of \ac{PV-DM} and \ac{PV-DBOW} are illustrated in Figure~\ref{fig:pv}. The PV-DBOW model is a Skip-Gram model where the input is a paragraph instead of a word. The \ac{PV-DBOW} is trained to predict a sample context given the input paragraph. In contrast, the \ac{PV-DM} model is trained to predict a word that is likely to occur in the input paragraph after the sample context. The \ac{PV-DM} model is a CBOW model extended with a paragraph in the input layer and a document embedding matrix. In the \ac{PV-DBOW} model, only paragraph embeddings are learned whereas in the \ac{PV-DM} model word embeddings and paragraph embeddings are learned, simultaneously. 


In Figure~\ref{fig:pv}, $p$ stands for the index of the input paragraph and $w_1,w_2,w_3,w_4$ represent the indices of the words in a contiguous sequence of words sampled from this paragraph. A sequence of size four is selected just for illustration purposes. Also, $D$ represents the paragraph embedding matrix and $E$ stands for the word embedding matrix. At the lowest layer, the input paragraph $p$ is mapped to its embedding by a lookup in the $D$ matrix. The hidden context representation is computed by summing the embeddings of the input words and paragraph, which is the same as in the \ac{CBOW} model.

Paragraph vector models are trained on unlabelled \emph{paragraph} collections. An embedding for each paragraph in the collection is learned at the end of training. The embedding for an unseen paragraph can be obtained by an additional inference stage. In the inference stage, $D$ is extended with columns for new paragraphs; $D$ is updated using gradient descent while other parameters of the model are kept fixed.

\citet{2014_Le_DistributedRepresentationsofSentencesandDocuments} assess vectors obtained by averaging \ac{PV-DM} and \ac{PV-DBOW} vectors on sentiment classification and snippet retrieval tasks. The snippet retrieval experiments are performed on a dataset of triplets created using snippets of the top 10 results retrieved by a search engine, for a set of 1 million queries. Each triplet is composed of two relevant snippets for a query and a randomly selected irrelevant snippet from the collection. Cosine similarity between paragraph vectors is shown to be an effective similarity metric for distinguishing similar snippets in such triplets. 
\citet{2014_Dai_DocumentEmbeddingwithParagraphVectors} show that paragraph vectors outperform the vector representations obtained by \ac{LDA}~\cite{blei2003latent}, average of word embeddings and tf-idf weighted one-hot vector representations, on a set of document triplets constructed with the same strategy in \cite{2014_Le_DistributedRepresentationsofSentencesandDocuments}, using Wikipedia and arXiv documents.


\section{Resources}
\label{sec:resources}

In this section, we present pointers to publicly available resources and tools which the reader would benefit for getting started with neural models, distributed representations, and information retrieval experiments with semantic matching.


\subsection{Introductory tutorials}

\citet{goldberg2015primer_tutorial} and \citet{cho2015_dist_tutorial} provide tutorials on getting started with neural networks from the natural language understanding perspective. \citet{goldberg2015primer_tutorial} covers details on training neural networks and a broader set of architectures including feed-forward networks, convolutional networks, recurrent networks and recursive networks. \citet{cho2015_dist_tutorial} focuses on language modeling and machine translation, sketches a clear picture of the encoder-decoder architectures, recurrent networks and attention modules~\cite{2014_NMT_With_Attention}.
 
\subsection{Word embeddings}

\paragraph{Corpora used}

Wikipedia and GigaWord5 are the corpora widely used for learning word embeddings. The latest Wikipedia dump can be obtained from Wikimedia.\footnote{\url{https://dumps.wikimedia.org/enwiki/latest/enwiki-latest-pages-articles.xml.bz2}} The GigaWord5 data set is accessible through the LDC.\footnote{\url{https://catalog.ldc.upenn.edu/LDC2011T07}} Several authors have learned embeddings from query logs~\cite{cai-learning-2016,2015_Sordoni_AHierarchicalRecurrentEncoder-DecoderforGenerativeContext-AwareQuerySuggestion,kanhabua2016learning}.

\paragraph{Pre-trained word embeddings}

It is possible to obtain pre-trained GloVe embeddings\footnote{\url{http://nlp.stanford.edu/projects/glove/}} and CBOW embeddings learned from Bing query logs~\cite{2016_Nalisnick_ImprovingDocumentRankingwithDualWordEmbeddings}. 

\paragraph{Learning word embeddings}

The source code for GloVe~\cite{2014_Pennington_Glove:GlobalVectorsforWordRepresentation}\footnote{\url{http://nlp.stanford.edu/projects/glove/}} and the models introduced in \citep{2015_Levy_Improving_Distributional_Similarity_With_Lessons_From_Word_Embeddings}\footnote{\url{https://bitbucket.org/omerlevy/hyperwords}} is publicly shared by the authors. Implementations of the Word2Vec and Paragraph Vector models are included in the \emph{gensim} library.\footnote{\url{https://radimrehurek.com/gensim/}}

\paragraph{Visualizing word embeddings}

The dimensionality reduction technique t-Distributed Stochastic Neighbor Embedding (t-SNE)~\cite{maaten2008visualizing} is commonly used for visualizing word embedding spaces and for trying to understand the structures learned by neural models.\footnote{\url{https://lvdmaaten.github.io/tsne/}}

\subsection{Test corpora used for retrieval experiments}

For retrieval experiments regarding neural models for semantic matching, the full range of CLEF, FIRE and TREC test collections has been used. Specifically, for evaluating neural models for semantic matching in an end-to-end web search task, TREC collections such as ClueWeb~\cite{guo-deep-2016}, .GOV~\cite{2015_Zuccon_Integrating_and_Evaluating_Neural_Word_Embeddings_In_Information_Retrieval}, GOV2~\cite{yang2016selective}, WT10G~\cite{2016_Roy_word_embeddings_for_query_expansion}, CLEF collections such as CLEF 2001--2003 Ad hoc~\cite{2015_Vulic_MonolingualandCross-LingualInformationRetrievalModelsBasedonBilingualWordEmbeddings} and CLEF 2003 English Ad hoc~\cite{2016_Masri_A_Comparison_of_Deep_Learning_Based_Query_Expansion_with_PRF_and_MI}, as well as logs such as the AOL log~\cite{cai-learning-2016,2015_Sordoni_AHierarchicalRecurrentEncoder-DecoderforGenerativeContext-AwareQuerySuggestion} and the MSN log~\cite{kanhabua2016learning} have been used.

\subsection{Implementing neural \ac{SC} models}

Theano,\footnote{\url{http://deeplearning.net/software/theano/}} TensorFlow\footnote{\url{https://www.tensorflow.org/}} and Torch\footnote{\url{http://torch.ch/}} are libraries that are widely used by the deep learning community for implementing neural network models. These libraries enable construction of neural network models from pre-defined high-level building blocks such as hidden units and layers. It is possible to define neural network models with different choices of architectures, non-linearity functions, etc.

GPU support and automatic differentiation~\cite{baydin2015automatic} are crucial features required for training neural networks. Theano and TensorFlow, enable performing matrix operations efficiently, in parallel, on GPU. Training neural networks with back-propagation requires computation of derivatives of the cost function with respect to every parameter of the network. Automatic differentiation in Theano and TensorFlow relieve the users from the effort on the manual derivation of derivatives of the objective function. These libraries compute the derivatives automatically given the definition of the neural network architecture and the cost function.  

The deep learning community has a strong tradition of sharing, in some version, the code used to produce the experimental results reported in the field's publications. The information retrieval is increasingly adopting this attitude too. Some authors of publications on neural models for semantic matching in web search have shared their code; see e.g., \cite{2015_Sordoni_AHierarchicalRecurrentEncoder-DecoderforGenerativeContext-AwareQuerySuggestion,vangysel-unsupervised-2016,vangysel-learning-2016,2015_Zuccon_Integrating_and_Evaluating_Neural_Word_Embeddings_In_Information_Retrieval}.\footnote{\url{https://github.com/ielab/adcs2015-NTLM}, \url{https://github.com/cvangysel/SERT},\url{https://github.com/cvangysel/SERT}}

\subsection{Experimenting with neural models}
Experimenting with neural networks is sometimes thought to be a bit of a ``dark art.'' Countless blog posts\footnote{See, e.g., \url{http://colah.github.io/}, \url{http://sebastianruder.com/#open}, \url{http://www.wildml.com/}, \url{http://neuralnetworksanddeeplearning.com/} to mention a few.} have been devoted to this as well as many posts on community question answering sites such as Stack Overflow\footnote{\url{http://stackoverflow.com/questions/tagged/deep-learning}} and Quora.\footnote{\url{https://www.quora.com/topic/Deep-Learning}} These are often valuable sources of advice, with source code snippets and illustrations.


\section{Taxonomy}
\label{sec:taxonomy}

As part of this introduction to neural models for semantic matching in web search, we provide a survey of related work. To organize the material, we use a simple taxonomy. We classify related work based on the \acfi{TTU}\acused{TTU} for which distributed representations are required, \emph{how} distributed representations are built, and the intended \emph{usage} of the distributed representations learned. We summarize these features in Table~\ref{tab:tax-features}. 

\begin{table}[h!]
\caption{Taxonomy features used for classifying publications n neural models for semantic matching.}
\label{tab:tax-features} 
\centering
\begin{tabular}{ p{.075\textwidth} p{.25\textwidth} p{.35\textwidth}  p{.125\textwidth} }
\toprule
\textbf{Label} & \textbf{Name} & \textbf{Explanation} & \textbf{Values} \\
\midrule
\ac{TTU}  & Target textual unit & Type of the textual unit type for which  distributed representations are computed & document \newline query \newline ad \\
\midrule
Usage  & Usage of distributed representations & How are \ac{TTU} representations used? & similarity \newline  feature \\
\midrule
How &  Method for building distributed representations & Does the method aggregate only external embeddings into \ac{TTU} representations? Is a neural semantic compositionality model learned?  &  aggregate \newline learn \\
\bottomrule
\end{tabular}
\end{table}

\subsection{TTU}
\label{sec:tax-ttu}
\noindent%
The set of \acp{TTU} depends on the target task. Distributed representations for queries are required for all the web search tasks covered in this survey. Representations for documents and ads are additionally required for document and ad retrieval tasks, respectively.

\subsection{Usage}
\label{sec:tax-use}

There are two choices of \emph{usage} observed in the work that we reviewed: 
\begin{description}
	\item[\bf Similarity] The representations are used to compute semantic similarity of \ac{TTU} pairs directly. 
	\item[\bf Feature] \ac{TTU} representations are used as additional or standalone features in existing supervised settings, mostly in learning to rank frameworks, for web search.    
\end{description}

\subsection{How}
\label{sec:tax-how}

The \emph{how} feature defines the method for building \ac{TTU} representations. The methods can be grouped in two main classes \emph{aggregate} and \emph{learn}, which we discuss below.

\subsubsection{Aggregate}
\label{sec:tax-aggregate}

The methods in this category rely on pre-trained word embeddings as external resources for distributed \ac{TTU} representations. Existing work can be split into two sub-categories depending on how the embeddings are utilized: 
\begin{description}
    \item[\textbf{Explicit}] Word embeddings are considered as building blocks for distributed \ac{TTU} representations. The works that follow this pattern treat a \ac{TTU} as \ac{BoEW} or a set of points in the word embedding space. The most common aggregation method is averaging or summing the vectors of the terms in the \ac{TTU}.
    \item[\textbf{Implicit}] Here, one utilizes the vector similarity in the embedding space in language modeling frameworks without explicit computation of similarity based on distributed representations for \acp{TTU}. For instance, \citet{2015_Zuccon_Integrating_and_Evaluating_Neural_Word_Embeddings_In_Information_Retrieval} compute translation probabilities of word pairs in a translation language model retrieval framework with cosine similarity of Skip-Gram vectors. 
\end{description} 

\subsubsection{Learn}
\label{sec:tax-learn}
This category covers work on learning neural semantic compositionality models for distributed representations. Three separate training objectives are observed in the reviewed work. Based on these objectives, we define the three sub-categories, namely  \emph{neural learn to match}, \emph{learn to predict context} and \emph{learn to generate context}, all of which are detailed below.

\paragraph{\textbf{Neural learning to match}}
\label{sec:learn-to-match}
\emph{Learning to match}~\citep{2013_Li_SemanticMatchinginSearch} is the problem of learning a matching function $f(x,y)$ that computes the similarity degree of two objects $x$ and $y$ from two different spaces $X$ and $Y$. Given training data $T$ composed of triples $(x,y,r)$, \emph{learning to match} is the optimization problem in Equation~\ref{eq:learntomatch}. There, $L$ denotes a loss function between the actual similarity score $r$ and the score predicted by the $f$ function: 
\begin{equation}
     \arg\min_{f \in F} \sum_{(x,y,r) \in T} L(r,f(x,y)).
     \label{eq:learntomatch}
\end{equation}
Neural learning to match models rely on distributed representations generated by neural networks for computing the similarity between objects. They are based on siamese neural networks~\cite{1993_bromly_siamese_network}, as illustrated in Figure~\ref{fig:siamese}, which consist of two identical sub-networks joined at their outputs in order to compute the similarity of the inputs. Input vector representations $x, y$ are mapped into distributed representations $h_x$, $h_y$ by the sub-network that we call \acfi{SCN}\acused{SCN} and the similarity score of the objects is computed by the similarity of the distributed representations. Usually, cosine similarity is used to compute the similarity vectors created by the \ac{SCN}, as given in Equations~\ref{eq:learn-to-match-scn} and~\ref{eq:learn-to-match-sim}:

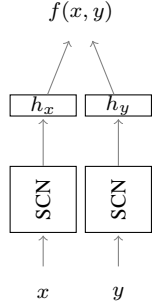
\begin{figure}[t]
\def\layersep{1.25cm}
\centering
\begin{tikzpicture}[shorten >=1pt,->,draw=black!50, node distance=4,transform shape,rotate=90]  
    \tikzstyle{every pin edge}=[<-,shorten <=1pt]
    \tikzstyle{neuron}=[circle,draw=black,minimum size=17pt,inner sep=0pt]
    \tikzstyle{vector}=[rectangle,draw=black,minimum width=8pt, minimum height=25pt,inner sep=0pt]
    \tikzstyle{subnet}=[rectangle,draw=black,minimum size=25pt,inner sep=0pt]
    \tikzstyle{justtex}=[rectangle,minimum size=20pt,inner sep=0pt]
    \tikzstyle{input neuron}=[neuron, fill=green!50];
    \tikzstyle{output neuron}=[neuron, fill=red!50];
    \tikzstyle{hidden neuron}=[neuron, fill=blue!50];
    \tikzstyle{annot} = [text width=4em, text centered]
    \tikzset{hoz/.style={rotate=-90}}   
    
   \path node[justtex,hoz,node distance=2cm] (Q) at (0,1) {$x$};
   \path node[justtex,hoz,node distance=2cm] (D) at (0,0) {$y$};
   \path node[justtex,hoz,node distance=2cm] (F) at (3*\layersep,0.5) {$f(x,y)$};

   \path node[subnet] (SC-Q) at (1*\layersep,1) {SCN};
   \path node[vector] (H-Q) at (2*\layersep,1) {};
   
   \path node[subnet] (SC-D) at (1*\layersep,0) {SCN};
   \path node[vector] (H-D) at (2*\layersep,0) {};
   
   \path node[justtex,hoz,node distance=2cm] () at (2*\layersep,1) {$h_x$};
   \path node[justtex,hoz,node distance=2cm] () at (2*\layersep,0) {$h_y$};

   \path (Q) edge (SC-Q);
   \path (SC-Q) edge (H-Q);
   \path (D) edge (SC-D);
   \path (SC-D) edge (H-D);   
   \path (H-D) edge (F);
   \path (H-Q) edge (F);

\end{tikzpicture}
\caption{A siamese network.}
\label{fig:siamese}
\end{figure} 
\begin{equation}\label{eq:learn-to-match-scn}
h_x = SCN(x), h_y = SCN(y)
\end{equation}
\begin{equation}\label{eq:learn-to-match-sim}
     f(x,y) = \frac{h_x \cdot h_y}{\left\vert{h_x}\right\vert \left\vert{h_y}\right\vert}.
\end{equation}
Learning to match models require similarity assessments for training. Since it is difficult to obtain large amounts of supervised data, click information in click-through logs are exploited to derive similarity assessments for query-document and query-ad pairs. If a pair $(x,y)$ is associated with clicks, the objects are accepted to be similar and dissimilar in the absence of clicks. In the case of query-query pairs, co-occurrence in a session is accepted as similarity signal that can be extracted from query logs. 

The training objective common to the models in this category is to minimise the negative log likelihood function in Equation~\ref{eq:learn-to-match-like}. We replace $x$ and $y$ with $q$ and $d$ to represent a query and a document object, respectively. The document object can be any textual unit that is needed to be matched against a query, such as a document, query or ad; $(q,d^+)$  denotes each (query-clicked document) pair extracted from logs and $D = \{d^+\} \cup D^-$ denotes the set of all documents in the collection. The likelihood of a document $d$ given a query $q$, is computed by Equation \ref{eq:learn-to-match-prob} with a softmax over similarity scores of distributed representations.

\begin{eqnarray}
     L &=& - \log \prod_{(q,d^+)}  P(d^+ \mid q)
\label{eq:learn-to-match-like}\\
     P(d\mid q) &=& \frac{\exp(\gamma f(q,d))}{\sum_{d' \in D} \exp(\gamma f(q,d')}
     \label{eq:learn-to-match-prob}
\end{eqnarray}
$L$ requires computation of a probability distribution over the entire document collection $D$ for each (query-clicked document) pair. Since this is computationally expensive, $D^-$ is approximated by a randomly selected small set of unclicked documents, similar to the softmax approximation methods for neural language models in Section \ref{subsec:eff-nlm}. In Figure~\ref{fig:siamese} the training strategy is illustrated with four randomly selected non-relevant documents. Let $d_1$ be a relevant document and ${d_2,d_3,d_4,d_5}$ non-relevant documents for the query $q$. The siamese network is applied to compute similarity scores for each pair $(q,d_i)$. 
 

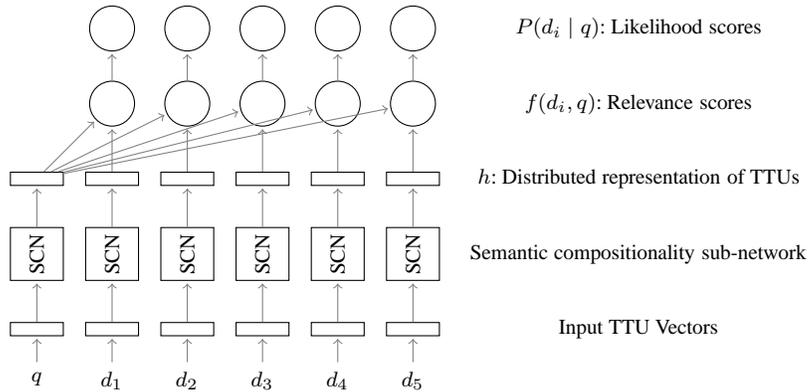
\begin{figure}[t]
\def\layersep{1cm}
\begin{tikzpicture}[shorten >=1pt,->,draw=black!50, node distance=\layersep,transform shape,rotate=90]  
    \tikzstyle{every pin edge}=[<-,shorten <=1pt]
    \tikzstyle{neuron}=[circle,draw=black,minimum size=17pt,inner sep=0pt]
    \tikzstyle{vector}=[rectangle,draw=black,minimum width=5pt, minimum height=20pt,inner sep=0pt]
    \tikzstyle{subnet}=[rectangle,draw=black,minimum size=20pt,inner sep=0pt]
    \tikzstyle{justtex}=[rectangle,minimum size=20pt,inner sep=0pt]
    \tikzstyle{input neuron}=[neuron, fill=green!50];
    \tikzstyle{output neuron}=[neuron, fill=red!50];
    \tikzstyle{hidden neuron}=[neuron, fill=blue!50];
    \tikzstyle{annot} = [text width=4em, text centered]
    \tikzset{hoz/.style={rotate=-90}}   
    
   \node[vector, pin=left:\rotatebox{-90}{\parbox[t][][r]{30mm}{\centering $q$}}] (Q) at (0,-1) {};
   \path node[subnet] (SC-Q) at (1*\layersep,-1) {SCN};
   \path node[vector] (H-Q) at (2*\layersep,-1) {};

   \path (Q) edge (SC-Q);
   \path (SC-Q) edge (H-Q);
   \path node[justtex,hoz] (L-Q) at (\layersep,-7) {};
   
      \path node[justtex,hoz,right of=I-5, node distance=2cm] (L-I) at (0,-7) {Input TTU Vectors};
   \path node[justtex,hoz,right of=SC-5, node distance=2cm] (L-SC) at (1*\layersep,-7) {Semantic compositionality sub-network};
   \path node[justtex,hoz,right of=H-5, node distance=2cm] (L-H) at (2*\layersep,-7) {$h$: Distributed representation of TTUs};
   \path node[justtex,hoz,right of=H-5, node distance=2cm] (L-R) at (3*\layersep,-7) {$f(d_i,q)$: Relevance scores};   
   \path node[justtex,hoz,right of=H-5, node distance=2cm] (L-R) at (4*\layersep,-7) {$P(d_i \mid q)$: Likelihood scores};  
      
    \foreach \name / \y in {1,...,5}
        \node[vector, pin=left:\rotatebox{-90}{\parbox[t][][r]{30mm}{\centering $d_\y$}}] (I-\name) at (0,-\y-1) {};
       
    \foreach \name / \y in {1,...,5}
        \path node[subnet] (SC-\name) at (1*\layersep,-\y-1) {SCN};

    \foreach \name / \y in {1,...,5}
        \path node[vector] (H-\name) at (2*\layersep,-\y-1) {};

    \foreach \name / \y in {1,...,5}
        \path node[neuron] (R-\name) at (3*\layersep,-\y-1) {};
    \foreach \name / \y in {1,...,5}
        \path node[neuron] (P-\name) at (4*\layersep,-\y-1) {};


    \foreach \source in {1,...,5}
            \path (I-\source) edge (SC-\source);
                  
    \foreach \source in {1,...,5}
            \path (SC-\source) edge (H-\source);
            
    \foreach \source in {1,...,5}
            \path (H-\source) edge (R-\source);
    
    \foreach \source in {1,...,5}
            \path (H-Q) edge (R-\source);
            
   \foreach \source in {1,...,5}
            \path (R-\source) edge (P-\source);
            


\end{tikzpicture}
\label{fig:dssm}
\caption{Architecture of the neural learn to match models adapted from \cite{deng-deep-2013}.}
\end{figure}

\paragraph{\textbf{Learn to predict context}}
\label{sec:learn-to-predict}
The success of word-based neural language models has motivated models for learning representations of larger textual units \citep{2014_Le_DistributedRepresentationsofSentencesandDocuments, 2016_Hill_LearningDistributedRepresentationsofSentencesfromUnlabelledData} from unlabelled data. As mentioned previously, neural language models are context-predicting \acp{DSM}. The \emph{learn to predict} models rely on an extension of the context-prediction idea to larger linguistic units, which is that \emph{similar textual units occur in similar contexts}. For instance, sentences in a paragraph and paragraphs in a document are semantically related. The organisation of textual units---of different granularity---in corpora created by human can be exploited to learn distributed representations. Contextual relationships of textual units are exploited to design training objectives similar to neural language models.

We refer to the models that rely on the extended distributional semantics principle to obtain distributed \ac{TTU} representations as \emph{learn to predict context} models in accordance with the \emph{context-predicting} label used for neural language models. \emph{Learn to predict context} models are neural network models trained using unlabelled data to maximize the likelihood of the context of a \ac{TTU}.  Among the models reviewed in \citep{2016_Hill_LearningDistributedRepresentationsofSentencesfromUnlabelledData}, Skip-thought Vector~\citep{2015_Kiros_Skip-ThoughtVectors} and \ac{PV}~\citep{2014_Le_DistributedRepresentationsofSentencesandDocuments} are successful representatives of the \emph{learn to predict context} idea. The training objective of the Skip-Thought Vector is to maximize the likelihood of the previous and the next sentences given an input sentence. The context of the sentence is defined as its neighboring sentences. Recall that details of the paragraph vector model are provided in Section~\ref{sec:background}. 

The main context of a textual unit is the words it contains. Containment should be seen as a form of co-occurrence and the content of a document is required to define its textual context. Two documents are similar if they contain similar words. Besides the content, temporal context is exploited in all studies in this category for defining \ac{TTU} contexts. For instance, the context of the query can be defined by the other queries in the session \citep{2015_Grbovic_Context-andContent-awareEmbeddingsforQueryRewritinginSponsoredSearch}. Finally, context of a document is defined by joining its content and the neighboring documents in a document stream in~\cite{2015_Djuric_HierarchicalNeuralLanguageModelsforJointRepresentationofStreamingDocumentsandtheirContent}. 

\paragraph{\textbf{Learn to generate context}}
\label{sec:learn-to-generate}

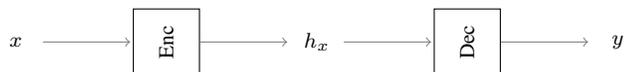
\begin{figure}[h!]
\def\layersep{1.25cm}
\centering
\begin{tikzpicture}[shorten >=1pt,->,draw=black!50, node distance=4,transform shape,rotate=90]  
    \tikzstyle{every pin edge}=[<-,shorten <=1pt]
    \tikzstyle{neuron}=[circle,draw=black,minimum size=17pt,inner sep=0pt]
    \tikzstyle{vector}=[rectangle,draw=black,minimum width=8pt, minimum height=25pt,inner sep=0pt]
    \tikzstyle{subnet}=[rectangle,draw=black,minimum size=25pt,inner sep=0pt]
    \tikzstyle{justtex}=[rectangle,minimum size=20pt,inner sep=0pt]
    \tikzstyle{input neuron}=[neuron, fill=green!50];
    \tikzstyle{output neuron}=[neuron, fill=red!50];
    \tikzstyle{hidden neuron}=[neuron, fill=blue!50];
    \tikzstyle{annot} = [text width=4em, text centered]
    \tikzset{hoz/.style={rotate=-90}}   

   \path node[justtex,hoz,node distance=2cm] (X) at (0,8) {$x$};
   \path node[justtex,hoz,node distance=2cm] (H_X) at (0,4) {$h_x$};
   \path node[justtex,hoz,node distance=2cm] (Y) at (0,0) {$y$};

   \path node[subnet] (D) at (0,2) {Dec};
   \path node[subnet] (E) at (0,6) {Enc};
  
   \path (X) edge (E);
   \path (E) edge (H_X);
   \path (H_X) edge (D);
   \path (D) edge (Y);

\end{tikzpicture}
\caption{Encoder decoder architecture.}
\label{fig:encdec}
\end{figure}

This category covers recent work in which a synthetic textual unit is generated based on the distributed representation of an input \ac{TTU}. Studies in this category are motivated by successful applications of RNNs, LSTM networks, and encoder-decoder architectures, illustrated in Figure~\ref{fig:encdec}, to sequence-to-sequence learning~\cite{Graves_Supervised_Sequence_Labelling_Book} tasks such as machine translation~\cite{2014_Cho_LearningPhraseRepresentationsusingRNNEncoder-DecoderforStatisticalMachineTranslation} and image captioning~\cite{Show-Attend-Tell-Image-Captioning}. In these applications, an input textual or visual object is encoded into a distributed representation with a neural network and a target sequence of words---a translation in the target language or a caption---is generated by a decoder network. 


\noindent%
\emph{Learn to predict context} models are focused on learning embeddings for a fixed collection of \acp{TTU}. In contrast, \emph{learn to generate context} models are designed to generate unseen synthetic textual units. In \emph{learn to predict context} models, an additional inference step is required for obtaining the embedding of an unobserved \ac{TTU}, as in the paragraph vector model. In the \emph{learn to generate context} models, distributed representations of unseen \acp{TTU} can be computed by forward propagation through the neural network model.

\subsection{Roadmap}

In Table~\ref{tab:high-tax}, we provide a classification of reviewed work with respect to the features. For the publications in the \emph{aggregate, implicit} category, the \emph{usage} feature is not applicable since an explicit distributed representation is not constructed. Besides, in \cite{2016_Lioma_Deep_Learning_Relevance}, which falls into \emph{learn to generate context} category, the generated document is not used in a quantitative experimental framework. These exceptions are indicated using the -- sign.


\begin{table}[h!]
\caption{Classification of reviewed work.}
\label{tab:high-tax} 
\centering
\begin{tabular}{ p{1.25cm}p{1.35cm}p{1cm}p{1cm}p{1.95cm}p{1.35cm}p{1cm}}
\toprule
\textbf{Task} & \textbf{TTU} & \textbf{Usage} & \textbf{How} & & \textbf{Works} & \textbf{Section}\\
\midrule
Doc. ret. & Query-Doc  & similarity & aggregate & explicit & \citep{2013_Clinchant_AggregatingContinuousWordEmbeddingsforInformationRetrieval,2015_Vulic_MonolingualandCross-LingualInformationRetrievalModelsBasedonBilingualWordEmbeddings,2016_Nalisnick_ImprovingDocumentRankingwithDualWordEmbeddings} & \ref{sec:adhoc-agg-exp}\\
 &  & &  & implicit & \citep{2015_Zuccon_Integrating_and_Evaluating_Neural_Word_Embeddings_In_Information_Retrieval,2015_Ganguly_WordEmbeddingBasedGeneralizedLanguageModelforInformationRetrieval} & \ref{sec:adhoc-agg-imp}\\\midrule 
Query exp. & Query & similarity & aggregate & explicit & \citep{2016_Masri_A_Comparison_of_Deep_Learning_Based_Query_Expansion_with_PRF_and_MI,2015_Zheng_LearningtoReweightTermswithDistributedRepresentations} & \ref{sec:adhoc-qexp-exp}\\
  &  &  -- &   & implicit & \citep{2016_Diaz_Query_Expansion_With_Locally_Trained_Word_Embeddings,2016_Zamani_Embedding_Based_Query_Language_Models} & \ref{sec:adhoc-qexp-imp}\\ 
\midrule
Doc. ret.  & Query-Doc & similarity & learn& learn-to-match & \citep{2013_Huang_LearningDeepStructuredSemanticModelsforWebSearchUsingClickthroughData, 2014_Shen_LearningSemanticRepresentationsUsingConvolutionalNeuralNetworksforWebSearch,2015_Palangi_DeepSentenceEmbeddingUsingtheLongShortTermMemoryNetwork:AnalysisandApplicationtoInformationRetrieval} & \ref{sec:adhoc-learn-to-match} \\
	 &  &  & & learn-to-predict & \citep{2015_Djuric_HierarchicalNeuralLanguageModelsforJointRepresentationofStreamingDocumentsandtheirContent,2016_Ai_Analysis_of_PV_for_IR} & \ref{sec:adhoc-learn-to-predict}\\
\midrule  
Doc. ret.  & Query  & -- & learn & learn-to-generate & \citep{2016_Lioma_Deep_Learning_Relevance} & \ref{sec:adhoc-learn-to-generate} \\\midrule
Query sug.  & Query & -- & aggregate & implicit &  \cite{cai-learning-2016} & \ref{sec:query-aggregate} \\\midrule
Query sug.  & Query & feature & learn & learn-to-match &  \cite{2015_Mitra_QueryAuto-CompletionforRarePrefixes, 2015_Mitra_ExploringSessionContextUsingDistributedRepresentationsofQueriesandReformulations} & \ref{sec:query-learn-to-match} \\
 &  & & learn & learn-to-generate &  \cite{2015_Sordoni_AHierarchicalRecurrentEncoder-DecoderforGenerativeContext-AwareQuerySuggestion} & \ref{sec:query-learn-to-generate} \\\midrule
Ad ret. & Query-Ad &  similarity & learn & learn-to-match & \cite{2015_Azimi_AdsKeywordRewritingUsingSearchEngineResults,2016_Zhai_AttentionBasedRecurrentNeuralNetworksforOnlineAdvertising, 2016_Zhai_DeepIntent} & \ref{sec:ads-learn-to-match} \\
\midrule
Ad ret. &  Query & similarity & learn  & learn-to-predict & \citep{2015_Grbovic_SearchRetargetingusingDirectedQueryEmbeddings,2015_Grbovic_Context-andContent-awareEmbeddingsforQueryRewritinginSponsoredSearch} & \ref{sec:ads-learn-to-predict}\\
\bottomrule
\end{tabular}
\end{table}

In Section~\ref{sec:adhoc}, \ref{sec:query}, and~\ref{sec:ads} below we survey work on neural models for semantic matching in document retrieval, query suggestion, and ad retrieval, respectively.

\section{Document retrieval}
\label{sec:adhoc}

In this section we survey work on neural models for semantic matching for document retrieval. We follow the \emph{how} feature as explained in Section~\ref{sec:taxonomy}  for organizing the material discussed. Since query expansion and query re-weighting aim to improve retrieval effectiveness by query analysis, publications on these tasks are also included in this section.

\subsection{Aggregate}

In this section, we present the works that rely on pre-trained word embeddings for document retrieval under the \emph{implicit} and \emph{explicit} categories. In each subsection, we mention the works on query expansion and query re-weighting, separately.

\subsubsection{Explicit}

\label{sec:adhoc-agg-exp}
This category includes publications in which query and document are viewed as \acf{BoEW}. Each word is associated with an embedding vector. Query-document relevance is computed using \ac{BoEW} representations for the query and document. Besides query-document matching, we present query expansion studies that rely on term similarities in the embedding space to find similar terms to query. 

\begin{table}[h!]
\caption{Summary of aggregation methods for document retrieval.}
\label{tab:aggregate} 
\centering
\begin{tabular}{ p{.5cm} p{1cm} p{2.25cm} p{2.5cm} p{3.5cm} }
\toprule
 & \textbf{NLM} &  \textbf{Query rep.} & \textbf{Document rep.} &  \textbf{Similarity func.} \\
\midrule
\cite{2015_Vulic_MonolingualandCross-LingualInformationRetrievalModelsBasedonBilingualWordEmbeddings} & Skip-gram & Sum of BoEW-IN & Sum over BoEW-IN  & Cosine similarity \\ \midrule
\cite{2016_Nalisnick_ImprovingDocumentRankingwithDualWordEmbeddings} & CBOW & BoEW-IN & BoEW-OUT & Aggregation of cosine similarities of accross all query-document pairs \\\midrule
\cite{2015_Kusner_FromWordEmbeddingsToDocumentDistances} & Skip-Gram & BoEW-IN & BoEW-IN & Word's Mover Distance (WMD)\\
\midrule 
\cite{2016_Roy_Representing_Documents_an_Queries_As_Sets_of_Words} & CBOW & BoEW-IN & Set of clusters in IN & Average inter-similarity of query to the set of centroids in the document \\
\midrule
\cite{2014_Palakodety_QueryTransformationsforResultMerging} & CBOW & Mean of BoEW-IN &  N/A  &  Cosine similarity \\\midrule
\cite{2016_Roy_word_embeddings_for_query_expansion} & Word2Vec & Mean of BoEW-IN &  N/A  &   The mean cosine similarity between expansion term and all the terms in
Q \\
\bottomrule
\end{tabular}
\end{table}

In Table \ref{tab:aggregate}, we provide a summary of the methods. The \ac{NLM} column specifies the model used for learning word embeddings, as mentioned by the authors. The second and third columns list the representations adopted for query and document, respectively. The IN and OUT suffixes in these columns indicate whether input or output embeddings are utilized. As mentioned in Section~\ref{subsec:nlm}, each \ac{NLM} comprises two matrices $E$ and $C$, of size $|V| \times d$. Here, $E$ is the matrix for the embedding layer which maps the input one-hot vectors to embedding vectors; $C$ is the weights matrix of the classifier layer. The $E$ matrix is referred as IN embeddings whereas the $C$ matrix as OUT embeddings, in accordance with the naming in \cite{2016_Nalisnick_ImprovingDocumentRankingwithDualWordEmbeddings}. The last two rows of Table~\ref{tab:aggregate} cover publications on query expansion and re-weighting. The similarity column for these publications indicates the functions utilized for computing term or term-query similarities. 

\citet{2016_Nalisnick_ImprovingDocumentRankingwithDualWordEmbeddings} point out an important feature of CBOW embeddings: the neighbors of a word represented with its IN embedding vector in the OUT space are topically similar words. And within the same embedding space, either IN or OUT, the neighbors are functionally similar words. Topically similar words are likely to co-occur in a local context whereas functionally similar words are likely to occur in similar contexts. For instance, for the term \emph{harvard}, \emph{faculty, alumni, graduate} are topically terms and \emph{yale, stanford, cornell} functionally similar terms. Motivated by this observation, \citet{2016_Nalisnick_ImprovingDocumentRankingwithDualWordEmbeddings} propose the \ac{DESM}. In the DESM a query is represented with a \ac{BoEW} in the IN space and the document is represented with a \ac{BoEW} in OUT space. Query-document similarity is computed by aggregating cosine similarities across all the query-document word pairs \citep{2016_Nalisnick_ImprovingDocumentRankingwithDualWordEmbeddings, 2016_Mitra_Adualembeddingspacemodelfordocumentranking}. \ac{DESM} is evaluated in a document ranking setting with both explicitly judged data sets and implicit feedback based data sets in \cite{2016_Mitra_Adualembeddingspacemodelfordocumentranking}. On explicitly judged of implicit feedback based test data sets, \ac{DESM} outperforms the BM25 and LSA baselines. However, when \ac{DESM} is used standalone for ranking documents from a large test collection, its performance is significantly lower than that of the LDA and BM25 baselines. \ac{DESM} is shown to be an effective signal for re-ranking the document set retrieved by a first-stage retrieval model. 

In the \emph{explicit} category, \citet{2015_Vulic_MonolingualandCross-LingualInformationRetrievalModelsBasedonBilingualWordEmbeddings} propose to construct query and document representations as a sum of word embeddings learned from a pseudo-bilingual document collection with a Skip-Gram model. Each document pair in a document-aligned translation corpus is mapped to a pseudo-bilingual document by merging source and target documents, removing sentence boundaries and shuffling the complete document. Owing to these shuffled pseudo-bilingual documents, the words from source and target language are mapped to the same embedding space. This approach for learning bilingual word embeddings is referred as \ac{BWESG}. Documents are ranked by the cosine similarity of their embedding vector to the query vector. The query-document representations are evaluated both on cross-lingual and mono-lingual retrieval tasks. For the monolingual experiments, ranking the proposed distributed representations outperforms ranking  LDA representations. Moreover, the word embedding-based representation is shown to bring more improvements in MAP scores when combined with a unigram language model, compared to LDA. Embedding-based representations yield higher MAP scores in cross-lingual retrieval experiments too. However, when combined with the unigram language model, LDA-based representations outperformed embedding-based representations in the EN-to-NL retrieval direction. When embed\-ding-based representations are integrated into the unigram language model combined with LDA representations, MAP scores in the EN-NL direction can be increased. Another lesson from this work is that an IDF-weighted sum of word embeddings yields representations that can produce higher MAP scores than the summation of word embeddings. Aligned English and Dutch Wikipedia articles together with EuroParl English-Dutch document pairs are used for learning word embeddings. Both monolingual and bilingual retrieval experiments are conducted using CLEF 2001--2003 collections. 

In \cite{2016_Roy_Representing_Documents_an_Queries_As_Sets_of_Words}, documents are modelled as a mixture distribution that generates the observed terms in the document. \citeauthor{2016_Roy_Representing_Documents_an_Queries_As_Sets_of_Words} estimate this distribution with k-means clustering of embeddings of the terms in the document. The likelihood of a query to be generated by the document is computed by the average inter-similarity of the set of query terms to the centroids of clusters in the document, in the word embedding space. For efficiency, the global vocabulary is clustered using Word2Vec embeddings in advance and document specific clusters are created by grouping the terms according to their global cluster ids. A centroid-based query likelihood function is evaluated in combination with language modeling with Jelinek-Mercer smoothing on the TREC 6-7-8 and TREC Robust data sets. A significant improvement is observed by the inclusion of word embedding-based query-likelihood function over the standalone \ac{LM} baseline.  

Although focused on document classification, it is worth mentioning the \textit{Word Mover's Distance} metric by~\citet{2015_Kusner_FromWordEmbeddingsToDocumentDistances} in the \emph{implicit} category since it proposes to integrate word embedding similarity into a document distance metric. \citeauthor{2015_Kusner_FromWordEmbeddingsToDocumentDistances} propose the document distance metric \acfi{WMD}\acused{WMD} inspired by the well-known transportation problem Earth Mover's Distance. Each document is represented as a set of points in the word embedding space. The \ac{WMD} of two documents is the minimum cumulative distance that all words in the first document need to travel to exactly match document the second document. Euclidean distance is used for computing word distances. The \ac{WMD} metric is evaluated with kNN document classification experiments on different data sets and different types of textual units such as recipes, news documents, medical abstracts, tweets and sentences. \ac{WMD} is reported to outperform the BoW, BM25, TF-IDF, \ac{LSI}, \ac{LDA}, Marginalized Stacked Denoising Autoencoder, and CCG Componential Counting Grid methods in terms of test error rate. 

A different explicit aggregation method based on \ac{BoEW} representations is the Fisher Kernel proposed in \cite{2013_Clinchant_AggregatingContinuousWordEmbeddingsforInformationRetrieval}. This work is based on \ac{LSI} embeddings, not on neural word embeddings yet it is worth mentioning here since it proposes a different aggregation method. The Fisher Kernel commonly used for constructing an image representation using local descriptors extracted from image patches is used to aggregate word embeddings into a fixed length document representation. The \emph{Fisher vector} representation is compared to document representations obtained \ac{LSI}, \ac{LDA} and \ac{PLSA} document representations on document clustering and ad-hoc \ac{IR} tasks. For ad-hoc retrieval experiments, CLEF'03, TREC 1 and 2 and ROBUST data sets are leveraged. In both tasks, the Fisher vector representation outperforms \ac{LSI}, \ac{LDA} and \ac{PLSA}. However, on the \ac{IR} task it is outperformed by the \ac{IR} baseline \ac{DFR}.

\paragraph{\textbf{Query expansion and re-weighting}}
\label{sec:adhoc-qexp-exp}
%
\citet{2016_Masri_A_Comparison_of_Deep_Learning_Based_Query_Expansion_with_PRF_and_MI} use Skip-Gram and CBOW embeddings for extracting most similar terms for a given query term. Neighbors of each query term in the embedding space are used to construct an expansion term set. Expansion terms are weighted in proportion to their frequency in the expanded query. The expansion method is evaluated on CLEF medical collections and compared to expansion with PRF and expansion with Mutual Information. Experiments show that expansion with word embeddings provides statistically significant improvements over the PRF method.  

The work on query expansion~\cite{2014_Palakodety_QueryTransformationsforResultMerging} based on word embeddings, although proposed for result merging in federated web search, can also be noted here since it introduces the idea of an \emph{importance vector} in the query re-weighting method in~\citep{2015_Zheng_LearningtoReweightTermswithDistributedRepresentations}. In \citep{2014_Palakodety_QueryTransformationsforResultMerging}, a query is represented by the mean vector of query term embeddings and k-nearest neighbors of the query vector are selected to expand the query. Expansion terms are re-weighted with respect to their distance to the query vector. A query term that is more distant to the query vector is assumed to contain more information and is assigned a higher weight. Query re-weighting is modelled as a linear regression problem from importance vectors to term weights in \cite{2015_Zheng_LearningtoReweightTermswithDistributedRepresentations}. The importance vector, which is the offset between the query vector and the term vector, is used as the feature vector for the query term. A linear regression model is trained using the ground-truth term weights computed by term recall weights which is the ratio of relevant documents that contain the query term $t$ to the total number of relevant documents to the query $q$. Weighted queries based on a learned regression model are evaluated in a retrieval setting and compared against the LM and BM-25 models using the data sets ROBUST04, WT10g, GOV2, ClueWeb09B. Two variants of the model DeepTR-BOW and DeepTR-SD are compared against unweighted queries, sequential dependency models and and weighted sequential dependency models. Statistically significant improvement is observed at high precision levels and throughout the rankings compared to the first two methods.

\citet{2016_Roy_word_embeddings_for_query_expansion} propose a set of query expansion methods, based on selecting $k$ nearest neighbors of the query terms in the word embedding space and ranking these terms with respect to their similarity to the whole query. The search space for neighbors either covers the entire vocabulary or is limited to terms in the top-ranking documents from an initial retrieval. All of the expansion methods yielded lower effectiveness scores than a statistical co-occurrence based feedback method, in experiments with the TREC 6, 7 and~8 and TREC Robust data set and the LM with Jelinek Mercer smoothing as the retrieval function. 

\subsubsection{Implicit}
\label{sec:adhoc-agg-imp}

\citet{2015_Zuccon_Integrating_and_Evaluating_Neural_Word_Embeddings_In_Information_Retrieval} compute translation probabilities of word pairs using the cosine similarity of Word2Vec vectors in a translation language model for retrieval. The translation model enriched by embedding similarity scores is called the \ac{NLTM}. \ac{NLTM} is shown to outperform the Dirichlet LM baseline and to be comparable to translation model with Mutual Information. 

Cosine similarity of Word2Vec embeddings is used in a similar way in the \ac{GLM}~\cite{2015_Ganguly_WordEmbeddingBasedGeneralizedLanguageModelforInformationRetrieval}. In the GLM, the probability of a query term $t$ to be generated by a document or the collection is modelled by generating an intermediate term $t'$ followed by a noisy channel model that transforms $t$ to $t'$. Cosine similarity of word embeddings is used for computing the transformation probabilities between the intermediate term and the actual term. The \ac{GLM} outperforms LM and LDA-smoothed LM baselines in terms of MAP score on the TREC 6, 7 and~8 and TREC Robust data sets. However, an LDA smoothed LM achieved higher recall scores. 

\paragraph{\textbf{Query expansion}}
\label{sec:adhoc-qexp-imp}
In \citep{2016_Diaz_Query_Expansion_With_Locally_Trained_Word_Embeddings,2016_Zamani_Embedding_Based_Query_Language_Models}, word embeddings are used for defining a new \ac{QLM}. A \ac{QLM} specifies a probability distribution $p(w\mid q)$ over all the terms in vocabulary. In query expansion with language modeling, the top $m$ terms $w$ that have the highest $p(w\mid q)$ value are selected as expansion terms \citep{2012_Carpineto_AQE_Survey}. \citet{2016_Diaz_Query_Expansion_With_Locally_Trained_Word_Embeddings}, propose a query expansion language model based on word embeddings learned from topic-constrained corpora. When word embeddings are learned from a topically-unconstrained corpora, they can be very general. Therefore, a query language model is defined based on word embeddings learned using a subset of documents sampled from a multinomial created by applying softmax on KL divergence scores of all documents in the corpus. The original query language model is interpolated with the query expansion language model which is defined by weights of terms computed by the $UU^Tq$ where $U$ is the $|V| \times d$ dimensional embedding matrix and $q$ is the $|V| \times 1$ dimensional term matrix. Locally-trained embeddings are compared against global embeddings on TREC12, Robust and ClueWeb 2009 Category B Web corpus.  Local embeddings are shown to yield higher NDCG@10 scores. Besides the local and global option, the authors also investigate the effect of using the target corpus versus an external corpus for learning word embeddings. A topically-constrained set of documents sampled from a general-purpose large corpus achieves the highest effectiveness scores. 
 
\citet{2016_Zamani_Embedding_Based_Query_Language_Models} propose two \acp{QLM} and an extended relevance model~\citep{2001_Lavrenko_Relevance_Based_Language_Models} based on word embeddings. In the first \ac{QLM}, $p(w\mid q)$ is computed by multiplying likelihood scores $p(w\mid t)$ given individual query terms whereas in the second \ac{QLM}, $p(w\mid q)$ is estimated by an additive model over $p(w\mid t)$ scores. The $p(w\mid t)$ scores are based on similarity of word embeddings. For measuring similarity of embeddings, a sigmoid function is applied on top of the cosine similarity in order to increase the discriminative ability. The reason for this choice is the observation that the cosine similarity of the 1000th closest neighbor to a word is not much lower than the similarity of the first closest neighbor. Besides the \acp{QLM}, a relevance model~\citep{2001_Lavrenko_Relevance_Based_Language_Models}, which computes a feedback query language model using embedding similarities in addition to term matching, is introduced. The proposed query language models are compared to Maximum Likelihood Estimation (MLE), \ac{GLM} \cite{2015_Ganguly_WordEmbeddingBasedGeneralizedLanguageModelforInformationRetrieval,2016_Masri_A_Comparison_of_Deep_Learning_Based_Query_Expansion_with_PRF_and_MI} on AP, Robust and GOV2 collections from TREC. The first \ac{QLM} is shown to be more effective in query expansion experiments. Regarding the PRF experiments, the embedding based relevance model combined with expansion using the first \ac{QLM} produces the highest scores. 

\subsubsection{Reflections on evaluation}

Evaluation of word embedding-based approaches has been performed on different retrieval data sets with different choices of corpora for learning word embeddings. It is difficult to derive conclusions about the comparative performance of the proposed approaches across publications due to the variety of experimental frameworks. In Table~\ref{tab:adhoc-eval}, we present an overview of the experimental frameworks adopted in these studies. The \emph{NLM} column specifies the neural language model and the \emph{Corpus for WE} column specifies the corpus used for learning word embeddings. The \emph{Retrieval corpus} column presents the document collection used in retrieval evaluation. Finally, the last column gives the methods against which the proposed method is compared.

\begin{table}[t]
\caption{Summary of evaluation frameworks of publications in the \emph{aggregate} category. (``Corpus for WE'' indicates the corpus that was used to create word embeddings.)}
\label{tab:adhoc-eval}
\centering
\begin{tabular}{p{0.5cm}p{1cm}p{2cm}p{2.5cm}p{3.5cm}}
\toprule
& \textbf{NLM} & \textbf{Corpus for WE} & \textbf{Retrieval corpus} & \textbf{Compared against} \\
\midrule
\citep{2015_Vulic_MonolingualandCross-LingualInformationRetrievalModelsBasedonBilingualWordEmbeddings}                              & Skip-Gram  & EuroParl corpus, Aligned English-Dutch Wikipedia Articles & CLEF En-Dutch Collection & Unigram LM \newline Unigram LM+LDA \newline  Google Translate + Unigram LM +LDA \\
\midrule
\citep{2015_Zuccon_Integrating_and_Evaluating_Neural_Word_Embeddings_In_Information_Retrieval}  &  Skipgram, CBOW &AP88-89, WSJ87-92, Wikipedia & TREC 6-7-8, DOTGOV, TREC Medical Records Track 2011-2012 & Dirichlet LM, \newline Translation Language Model with Mutual Information \\
\midrule
\citep{2015_Ganguly_WordEmbeddingBasedGeneralizedLanguageModelforInformationRetrieval} & CBOW & TREC Document Collection & TREC 6-7-8  & LM, LDA smoothed LM\\
\midrule
\citep{2016_Mitra_Adualembeddingspacemodelfordocumentranking}  & CBOW &  Bing logs &  Document collection from Bing logs                                &   LSA, BM25\\
\midrule
\citep{2016_Zamani_Embedding_Based_Query_Language_Models}  & Glove
& Wikipedia 2014 + Gigawords & AP, Robust and GOV2 from TREC & \citep{2016_Masri_A_Comparison_of_Deep_Learning_Based_Query_Expansion_with_PRF_and_MI,2015_Ganguly_WordEmbeddingBasedGeneralizedLanguageModelforInformationRetrieval} \\
\midrule
\citep{2015_Zheng_LearningtoReweightTermswithDistributedRepresentations}  &  CBOW & Google News / ClueWeb09B / Retrieval Corpus   & ROBUST04 \newline  WT10g \newline  GOV2 \newline ClueWeb09B & Sequential dependency Model, Unweighted Queries\\
\midrule
\citep{2016_Masri_A_Comparison_of_Deep_Learning_Based_Query_Expansion_with_PRF_and_MI} & CBOW
& Image2009, Case2011 and Case2012 from CLEF & Image2011, Image 2012, Image 2012 and Case2011 from CLEF & Expansion with PRF and Mutual Information \\
\midrule
\citep{2016_Diaz_Query_Expansion_With_Locally_Trained_Word_Embeddings} & CBOW &  Wikipedia / Gigawords / Retrieval Corpus & TREC12, Robust and ClueWeb 2009 Category B & QL\\
\midrule
\citep{2016_Roy_Representing_Documents_an_Queries_As_Sets_of_Words} &  CBOW & Retrieval Corpus &  TREC 6-7-8 and Robust  & LM with Jelinek Mercer Smoothing \\
\bottomrule
\end{tabular}
\end{table}

Only a recent study \cite{2016_Zamani_Embedding_Based_Query_Language_Models} presents comparison of the embedding based query language model~\cite{2016_Zamani_Embedding_Based_Query_Language_Models} to the GLM~\cite{2015_Ganguly_WordEmbeddingBasedGeneralizedLanguageModelforInformationRetrieval} and NLTM~\cite{2015_Zuccon_Integrating_and_Evaluating_Neural_Word_Embeddings_In_Information_Retrieval}. Using term similarities computed by word embeddings in a query language model or a document language model is shown to improve retrieval effectiveness scores over the standard retrieval baseline. 

Among the models that fall within the \emph{aggregate} category, directly using word embeddings provides consistent gains in \cite{2015_Ganguly_WordEmbeddingBasedGeneralizedLanguageModelforInformationRetrieval} but not in \cite{2015_Zuccon_Integrating_and_Evaluating_Neural_Word_Embeddings_In_Information_Retrieval,2016_Diaz_Query_Expansion_With_Locally_Trained_Word_Embeddings,2016_Zamani_Embedding_Based_Query_Language_Models,2016_Zamani_Estimating_Embedding_Vectors_For_Queries,
	2016_Nalisnick_ImprovingDocumentRankingwithDualWordEmbeddings}. In \cite{2015_Zuccon_Integrating_and_Evaluating_Neural_Word_Embeddings_In_Information_Retrieval}, word embedding similarity achieves comparable effectiveness to mutual information-based term similarity. For query-document similarity, \citet{2016_Nalisnick_ImprovingDocumentRankingwithDualWordEmbeddings} point out that utilising relations between the IN and OUT embedding spaces learned by CBOW yields a more effective similarity function for query-document pairs. \citet{2016_Diaz_Query_Expansion_With_Locally_Trained_Word_Embeddings} propose to learn word embeddings from a topically constrained corpora since the word embeddings learned from an unconstrained corpus are found to be too general. \citet{2016_Zamani_Embedding_Based_Query_Language_Models,2016_Zamani_Estimating_Embedding_Vectors_For_Queries} apply a sigmoid function on the cosine similarity scores in order to increase the discriminative power.

The influence of choice of hyper-parameters used to learn word embeddings on retrieval effectiveness is not systematically analyzed in all of the publications in the \emph{aggregate} category. 

\paragraph{Neural language model}
The majority of the studies rely on CBOW and Skip-Gram models from the Word2Vec framework. Surprisingly, the Glove and Word2Vec models are not compared in any of the studies. The effect of the model choice is only investigated in \citep{2015_Zuccon_Integrating_and_Evaluating_Neural_Word_Embeddings_In_Information_Retrieval}. A significant difference is not observed between Skip-gram and CBOW in terms of their effect on retrieval effectiveness in a translation language model. A noted observation is that Skip-gram model is able to yield higher effectiveness scores with lower embedding dimension and context-window size.  

\paragraph{Corpus}
Embeddings learned from a general-purpose corpus like Wikipedia (\emph{general purpose embeddings}) and  embeddings learned from the retrieval corpus itself \emph{(corpus-specific word embeddings)} are compared in \citep{2015_Zuccon_Integrating_and_Evaluating_Neural_Word_Embeddings_In_Information_Retrieval,2015_Zheng_LearningtoReweightTermswithDistributedRepresentations}. A notable difference is not observed between corpus-specific embeddings and general-purpose word embeddings when used for query-reweighting~\cite{2015_Zheng_LearningtoReweightTermswithDistributedRepresentations} or in a translation language model for computing term similarities~\cite{2015_Zuccon_Integrating_and_Evaluating_Neural_Word_Embeddings_In_Information_Retrieval}. However, \citet{2016_Diaz_Query_Expansion_With_Locally_Trained_Word_Embeddings} highlight that query expansion with word embeddings learned from topic-constrained collection of documents, yield higher effectiveness scores compared to embeddings learned from a general-purpose corpora.
 
\paragraph{Embedding dimension and context-window size}
\citet{2015_Zuccon_Integrating_and_Evaluating_Neural_Word_Embeddings_In_Information_Retrieval} report that embeddings learned using Skip-Gram and CBOW models are robust to different choices of embedding dimensionality and context window size. A similar observation about effect of embedding size and context-size on retrieval effectiveness is shared by \citet{2015_Vulic_MonolingualandCross-LingualInformationRetrievalModelsBasedonBilingualWordEmbeddings}. The retrieval effectiveness is found to be stable with embedding dimensions greater than 300 and context window size greater than 30 in~\citep{2015_Vulic_MonolingualandCross-LingualInformationRetrievalModelsBasedonBilingualWordEmbeddings}.

\subsection{Learn}

In the \emph{learn} category (see Section~\ref{sec:tax-learn}) we consider three sub-categories, which we discuss next.

\subsubsection{Learn to match}
\label{sec:adhoc-learn-to-match}

In Table~\ref{tab:adhoc-learn-to-match}, neural learn to match models for document retrieval are summarized with the type of \ac{TTU} pairs and the type of the neural network used as the \ac{SCN}.

The \ac{DSSM}~\citep{2013_Huang_LearningDeepStructuredSemanticModelsforWebSearchUsingClickthroughData} is the earliest neural learn to match model. The \ac{SCN} of the \ac{DSSM} model is composed of a word hashing layer and a deep neural network with three non-linear layers. \ac{DSSM} takes the term vector of a document as the input vectors for query and documents. The word hashing layer converts input term vectors into a trigram vector. The vocabulary size is reduced from 500K to 30K by replacing each term with its letter trigrams. For instance, the word \emph{vector} is mapped to \emph{\{.ve, vec, ect, cto, tor ,or.\}} where the dot sign is used as the start and end character. The trigram vector is propagated through three non-linear neural networks.

\begin{table}[t]
\centering
\caption{Neural learn to match models for document retrieval in web search.}
\label{tab:adhoc-learn-to-match}
\begin{tabular}{p{2cm}p{2.5cm}p{5cm}}
\toprule
\textbf{Study} & \textbf{Pair Type} & \textbf{SCN}  \\
\midrule
DSSM~\cite{2013_Huang_LearningDeepStructuredSemanticModelsforWebSearchUsingClickthroughData} & Query-Document  & Feed Forward Deep Neural Network (DNN)  \\
\midrule
CLSM~\cite{2014_Shen_ALatentSemanticModelwithConvolutional-PoolingStructureforInformationRetrieval,2014_Shen_LearningSemanticRepresentationsUsingConvolutionalNeuralNetworksforWebSearch} & Query-Document (Title)  & Convolutional Neural Network (CNN)  \\
\midrule
LSTM-DSSM\cite{2015_Palangi_DeepSentenceEmbeddingUsingtheLongShortTermMemoryNetwork:AnalysisandApplicationtoInformationRetrieval} & Query-Document Title  & Long Short Term Memory (LSTM) Network  \\
\bottomrule
\end{tabular}
\end{table}

The \ac{CLSM}~\citep{2014_Shen_ALatentSemanticModelwithConvolutional-PoolingStructureforInformationRetrieval,2014_Shen_LearningSemanticRepresentationsUsingConvolutionalNeuralNetworksforWebSearch} and LSTM-DSSM~\citep{2014_Palangi_SemanticModellingwithLong-Short-TermMemoryforInformationRetrieval,2016_Palangi_DeepSentenceEmbeddingUsingLongShort-TermMemoryNetworks:AnalysisandApplicationtoInformationRetrieval} differ from \ac{DSSM} in the input representations and architecture of the SCN component. \ac{DSSM} takes a term vector of the textual unit and treats it as a bag of words. In contrast, \ac{CLSM} and LSTM-DSSM take a sequence of one-hot vectors of terms and treat the \acp{TTU} as a sequence of words. \ac{CLSM} includes a convolutional neural network and LSTM-DSSM includes an LSTM network as \ac{SCN}. Note that the word hashing layer is common to all three models.

\ac{DSSM}, \ac{CLSM} and LSTM-DSSM are evaluated on large-scale real data sets from Bing in  \citep{2013_Huang_LearningDeepStructuredSemanticModelsforWebSearchUsingClickthroughData,2014_Shen_ALatentSemanticModelwithConvolutional-PoolingStructureforInformationRetrieval,2014_Shen_LearningSemanticRepresentationsUsingConvolutionalNeuralNetworksforWebSearch}. In \cite{2013_Huang_LearningDeepStructuredSemanticModelsforWebSearchUsingClickthroughData}, \ac{DSSM} is shown to outperform the Word Translation Model, BM25, TF-IDF and Bilingual Topic Models with posterior regularization in terms of NDCG at cutoff values 1, 3 and 10. \ac{CLSM} is shown to outperform \ac{DSSM} in \cite{2014_Shen_LearningSemanticRepresentationsUsingConvolutionalNeuralNetworksforWebSearch}. Finally, LSTM-DSSM outperforms \ac{CLSM} in \cite{2016_Palangi_DeepSentenceEmbeddingUsingLongShort-TermMemoryNetworks:AnalysisandApplicationtoInformationRetrieval} when document titles are used instead of full document content. When document titles are used instead of the full document content, higher NDCG scores are achieved by\cite{2014_Shen_LearningSemanticRepresentationsUsingConvolutionalNeuralNetworksforWebSearch}. For computational reasons, LSTM-DSSM is evaluated only with document titles \cite{2016_Palangi_DeepSentenceEmbeddingUsingLongShort-TermMemoryNetworks:AnalysisandApplicationtoInformationRetrieval}. 

Another interesting publication that follows \ac{DSSM} is by 
\citet{2015_Liu_RepresentationLearningUsingMulti-TaskDeepNeuralNetworksforSemanticClassificationandInformationRetrieval} who propose a neural model with multi-task objectives. A model that integrates a deep neural network for query classification and the \ac{DSSM} model for web document ranking via shared layers is proposed. The word hashing layer and semantic representation layer of \ac{DSSM} are shared between the two models. The integrated network comprises separate task-specific semantic representation layers and output layers for two different tasks. A separate cost function is defined for each task. During training, in each iteration, a task is selected randomly and the model is updated only according to the selected cost function. The proposed model is evaluated on large scale commercial search logs. Experimental results show improvements by the integrated model over both standalone deep neural networks for query classification and a standalone \ac{DSSM} for web search ranking.

\citet{2014_Li_Deeplearningpoweredin-sessioncontextualrankingusingclickthroughdata} utilize distributed representations produced by \ac{DSSM} and \ac{CLSM} in order to re-rank documents based on in-session contextual information. Similarity of query-query and query-document pairs extracted from the session context is computed using \ac{DSSM} and \ac{CLSM} vectors. These similarity scores are included as additional features to represent session context in a context-aware learning to rank framework. 

\subsubsection{Learn to predict}
\label{sec:adhoc-learn-to-predict}

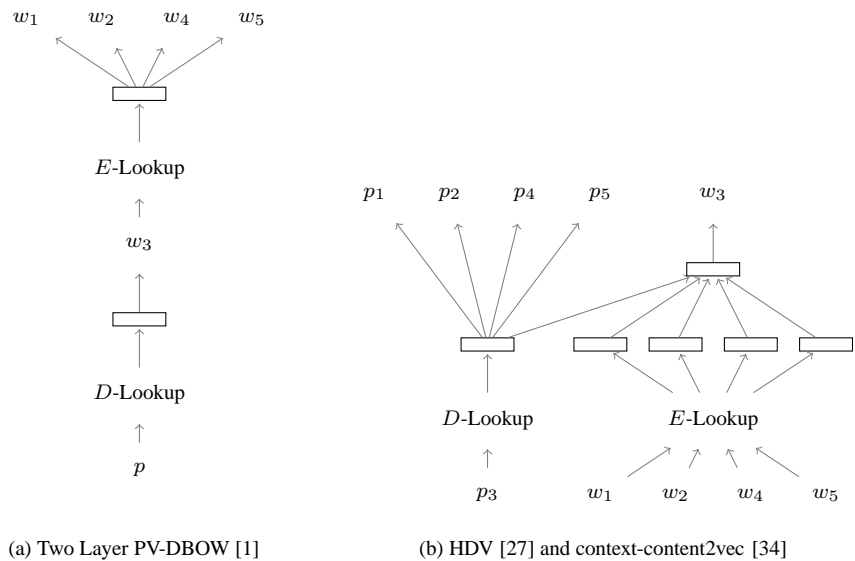
\begin{figure}[t]
  \centering
  \begin{minipage}[b]{0.3\textwidth}
    \centering

\def\layersep{1cm}
\begin{tikzpicture}[shorten >=1pt,->,draw=black!50, node distance=\layersep,transform shape,rotate=90]  
    \tikzstyle{every pin edge}=[<-,shorten <=1pt]
    \tikzstyle{neuron}=[circle,draw=black,minimum size=17pt,inner sep=0pt]
    \tikzstyle{vector}=[rectangle,draw=black,minimum width=5pt, minimum height=20pt,inner sep=0pt]
    \tikzstyle{layer}=[rectangle,draw=black,minimum width=5pt, minimum height=50pt,inner sep=0pt]
    \tikzstyle{subnet}=[rectangle,draw=black,minimum size=20pt,inner sep=0pt]
    \tikzstyle{justtex}=[rectangle,minimum size=20pt,inner sep=0pt]
    \tikzstyle{input neuron}=[neuron, fill=green!50];
    \tikzstyle{output neuron}=[neuron, fill=red!50];
    \tikzstyle{hidden neuron}=[neuron, fill=blue!50];
    \tikzstyle{annot} = [text width=4em, text centered]
    \tikzset{hoz/.style={rotate=-90}}   
   
   \path node[justtex,hoz,node distance=2cm] (O-1) at (6*\layersep,3.5) {$w_1$};
   \path node[justtex,hoz,node distance=2cm] (O-2) at (6*\layersep,2.5) {$w_2$};
   \path node[justtex,hoz,node distance=2cm] (O-3) at (6*\layersep,1.5) {$w_4$};
   \path node[justtex,hoz,node distance=2cm] (O-4) at (6*\layersep,0.5) {$w_5$};
   \path node[justtex,hoz,node distance=2cm] (W) at (3,2) {$w_3$};

   \path node[justtex,hoz,node distance=2cm] (D) at (0*\layersep, 2) {$p$};
   \path node[justtex,hoz,node distance=2cm] (PEM) at (1*\layersep, 2) {$D$-Lookup};
   \path node[justtex,hoz,node distance=2cm] (WEM) at (4*\layersep, 2) {$E$-Lookup};

   \path node[vector] (E-P) at (2,2) {};
   \path node[vector] (E-W) at (5,2) {};
   \path (D) edge (PEM);
   \path (PEM) edge (E-P);
   \path (W) edge (WEM);
   \path (E-P) edge (W);
   \path (WEM) edge (E-W);
   \foreach \source in {1,...,4}
       \path (E-W) edge (O-\source);   
\end{tikzpicture}
    \subcaption{Two Layer PV-DBOW~\cite{2016_Ai_Analysis_of_PV_for_IR}}
    \label{fig:pv-dbow2}
  \end{minipage}
  \hfill
  \begin{minipage}[b]{0.65\textwidth}
    \centering
    \def\layersep{1cm}
\begin{tikzpicture}[shorten >=1pt,->,draw=black!50, node distance=\layersep,transform shape,rotate=90]  
    \tikzstyle{every pin edge}=[<-,shorten <=1pt]
    \tikzstyle{neuron}=[circle,draw=black,minimum size=17pt,inner sep=0pt]
    \tikzstyle{vector}=[rectangle,draw=black,minimum width=5pt, minimum height=20pt,inner sep=0pt]
    \tikzstyle{layer}=[rectangle,draw=black,minimum width=5pt, minimum height=50pt,inner sep=0pt]
    \tikzstyle{subnet}=[rectangle,draw=black,minimum size=20pt,inner sep=0pt]
    \tikzstyle{justtex}=[rectangle,minimum size=20pt,inner sep=0pt]
    \tikzstyle{input neuron}=[neuron, fill=green!50];
    \tikzstyle{output neuron}=[neuron, fill=red!50];
    \tikzstyle{hidden neuron}=[neuron, fill=blue!50];
    \tikzstyle{annot} = [text width=4em, text centered]
    \tikzset{hoz/.style={rotate=-90}}   
    
   \path node[justtex,hoz] (I-1) at (0,3.5){$w_1$};
   \path node[justtex,hoz] (I-2) at (0,2.5){$w_2$};
   \path node[justtex,hoz] (I-3) at (0,1.5){$w_4$};
   \path node[justtex,hoz] (I-4) at (0,0.5){$w_5$};
   \path node[justtex,hoz] (O) at (4, 2) {$w_3$};
      
   \path node[justtex,hoz] (P-1) at (4,6.5) {$p_1$};
   \path node[justtex,hoz] (P-2) at (4,5.5) {$p_2$};
   \path node[justtex,hoz] (P-4) at (4,4.5) {$p_4$};
   \path node[justtex,hoz] (P-5) at (4,3.5) {$p_5$};

   \path node[justtex,hoz] (WEM) at (1, 2) {$E$-Lookup};
   \path node[justtex,hoz] (D) at (0, 5) {$p_3$};
   \path node[justtex,hoz] (PEM) at (1, 5) {$D$-Lookup};
   
   \path (D) edge (PEM);
   \path node[vector] (E-P) at (2,5) {};
   \path (PEM) edge (E-P);

   \foreach \source in {1,...,4}
            \path (I-\source) edge (E-L);

   \path node[vector] (E-1) at (2,3.5) {};
   \path node[vector] (E-2) at (2,2.5) {};
   \path node[vector] (E-3) at (2,1.5) {};
   \path node[vector] (E-4) at (2,0.5) {};

   \foreach \source in {1,...,4}
            \path (E-L) edge (E-\source);
   
   \path node[vector] (H-V) at (3,2) {};
   
   \path (E-P) edge (H-V);
   \path (E-P) edge (P-1);
   \path (E-P) edge (P-2);
   \path (E-P) edge (P-4);
   \path (E-P) edge (P-5);

   \foreach \source in {1,...,4}
            \path (E-\source) edge (H-V);
    \path (H-V) edge (O);
\end{tikzpicture}
    \subcaption{HDV~\cite{2015_Djuric_HierarchicalNeuralLanguageModelsforJointRepresentationofStreamingDocumentsandtheirContent} and context-content2vec~\cite{2015_Grbovic_Context-andContent-awareEmbeddingsforQueryRewritinginSponsoredSearch}}
    \label{fig:hdv}
  \end{minipage}
  \caption{Learn to predict context models for document retrieval.}
\end{figure}

\citet{2016_Ai_Improving_Language_Estimation_with_the_PV_Model_for_Ad-Hoc_Retrieval,2016_Ai_Analysis_of_PV_for_IR} investigate the use of the \ac{PV-DBOW} model as a document language model for retrieval. Three shortcomings of the \ac{PV-DBOW} model are identified and an extended paragraph vector model is proposed with remedies for these shortcomings. First, the \ac{PV-DBOW} model is found to be biased towards short documents due to overfitting in training and the training objective is updated with L2 regularization. Secondly, the \ac{PV-DBOW} model trained with NEG implicitly weights terms with respect to Inverse Corpus Frequencies (ICF) which has been shown to be inferior to Inverse Document Frequency (IDF) in \cite{robertson2004understandingTfidf}. A document frequency based negative sampling strategy, which converts the problem into factorization of a shifted TF-IDF matrix, is adopted. Thirdly, the two layer \ac{PV-DBOW} architecture depicted in Figure~\ref{fig:pv-dbow2} is introduced since word substitution relations, such as the relation in \emph{car-vehicle}, \emph{underground-subway} pairs, are not captured by \ac{PV-DBOW}. The \ac{EPV} is evaluated in re-ranking the set of top 2,000 documents retrieved by the QL retrieval function. Document language models based on \ac{EPV} and LDA are compared on TREC Robust04 and GOV2 data sets. An \ac{EPV}-based model yields higher effectiveness scores than the LDA-based model.

The \ac{HDV}~\citep{2015_Djuric_HierarchicalNeuralLanguageModelsforJointRepresentationofStreamingDocumentsandtheirContent} model extends the \ac{PV-DM} model to predict not only words in a document but also its temporal neighbors in a document stream. The architecture of this model is depicted in Figure~\ref{fig:hdv} with a word context and document context size of five. There, $w_1,w_2,w_3,w_4,w_5$ represent a sample context of words from the input paragraph; $p_1,p_2,p_3,p_4,p_5$ represent a set of documents that occur in the same context. 


In \ac{HDV}, the contents of the documents in the temporal context also contribute to the document representation. Similar to the \ac{PV-DM}, the words and documents are mapped to $d$-dimensional embedding vectors. \citet{2015_Djuric_HierarchicalNeuralLanguageModelsforJointRepresentationofStreamingDocumentsandtheirContent} point out that words and documents are embedded in the same space and this makes the model useful for both recommendation and retrieval tasks including document retrieval, document recommendation, document tag recommendation and keyword suggestion. Given a keyword, titles of similar documents in the embedding space are presented to give an idea of the effectiveness of the model on the ad-hoc retrieval task. However, a quantitative evaluation is not provided for the document retrieval and keyword suggestion tasks. 

\subsubsection{Learn to generate}
\label{sec:adhoc-learn-to-generate}

\citet{2016_Lioma_Deep_Learning_Relevance} ask whether it is possible to generate relevant documents given a query. A character level LSTM network is optimized to generate a synthetic document. The network is fed with a sequence of words constructed by concatenating the query and context windows around query terms in all relevant documents for the query. For each query, a separate model is generated and a synthetic document is generated for the same query with the learned model. The synthetic document was evaluated in a crowdsourcing setting. Users are provided with four word clouds that belong to three known relevant documents and the synthetic document. Each word cloud is built by selection of top frequent terms from the document. Users are asked to select the most relevant word cloud. Author report that the word cloud of the synthetic document ranked the first or second for most of the queries. Experiments were performed on the TREC Disks 4, 5 test collection with title-only queries from TREC~6, 7, 8. 

\subsection{Reflections on evaluation}
\label{sec:emb-effect}

To the best of our our knowledge, there are no experiments in the literature that comparison the \emph{aggregate} and \emph{learn} methods or that compare different training objectives for learning neural \ac{SC} models. Moreover, the models from \citep{2013_Huang_LearningDeepStructuredSemanticModelsforWebSearchUsingClickthroughData, 2014_Shen_LearningSemanticRepresentationsUsingConvolutionalNeuralNetworksforWebSearch,2015_Palangi_DeepSentenceEmbeddingUsingtheLongShortTermMemoryNetwork:AnalysisandApplicationtoInformationRetrieval} in the \emph{neural learn to match} category are evaluated on data sets derived from a commercial search engine and no quantitative evaluation is provided for the \emph{learn to predict} model \ac{HDV}~\cite{2015_Djuric_HierarchicalNeuralLanguageModelsforJointRepresentationofStreamingDocumentsandtheirContent}.


\section{Query suggestion}
\label{sec:query}

Next we turn to neural models for semantic matching for query suggestions. As explained in Section~\ref{sec:taxonomy}, for the query suggestion task the literature has publications in the \emph{aggregate} and \emph{learn} (\emph{learn to match} and \emph{learn to generate context}) categories.

\subsection{Aggregate}

\label{sec:query-aggregate}

The work by \citet{cai-learning-2016} on query auto-completion, introduces semantic features computed using Skip-Gram embeddings, for learning to rank query auto-completion candidates.  Query similarity computed by sum and maximum of the embedding similarity of term-pairs from queries are used as two separate features. The maximal embedding similarity of term pairs is found to be the most important feature in a diverse feature set including popularity-based features, a lexical similarity and another semantic similarity feature based on co-occurrence of term pairs in sessions. 

\subsection{Learn}
\label{sec:query-learn}

\subsubsection{Learn to match}
\label{sec:query-learn-to-match}

The \ac{CLSM}~\citep{2014_Shen_LearningSemanticRepresentationsUsingConvolutionalNeuralNetworksforWebSearch} has been used to learn distributed representations of queries~\cite{2015_Mitra_QueryAuto-CompletionforRarePrefixes} and query reformulations~\cite{2015_Mitra_ExploringSessionContextUsingDistributedRepresentationsofQueriesandReformulations}. In both studies, \ac{CLSM} representations are used to build additional features in an existing learning to rank framework for query auto-completion.

\citet{2015_Mitra_QueryAuto-CompletionforRarePrefixes} train a \ac{CLSM} model on query prefix-suffix pairs extracted from query logs by segmenting each query at every word boundary. The queries that start with the last word of the prefix issued by the user are picked from query logs. A candidate suggestion query is formed by appending the suffix from each such query to the prefix. A learning to rank model is trained using n-gram features and \ac{CLSM} based features, for ranking candidate queries. \ac{CLSM} representations are found to improve MMR scores when used together with n-gram features in experiments with AOL query logs and Bing logs. However, n-gram features are shown to provide a bigger improvement than \ac{CLSM} features.
 
In~\citep{2015_Mitra_ExploringSessionContextUsingDistributedRepresentationsofQueriesandReformulations}, a \ac{CLSM} model is trained on query pairs that are observed in succession in search logs. This work provides an analysis of \ac{CLSM} vectors for queries  similar to the word embedding space analysis in \citep{2013_Mikolov_LinguisticRegularitiesinContinuousSpaceWordRepresentations}.  \citet{2015_Mitra_ExploringSessionContextUsingDistributedRepresentationsofQueriesandReformulations} found out that offsets between CLSM query vectors can represent intent transition patterns. To illustrate, the nearest neighbour query of the  vector computed by $\mathit{vector}(\mathit{university~of~washington}) - \mathit{vector}(\mathit{seattle}) +\mathit{vector}(\mathit{chicago}€)$ is found to be $\mathit{vector}(\mathit{university~of~chicago})$. Besides, the offset of the vectors for $\mathit{university~of~washington}$ and $\mathit{seattle}$ is similar to the offset of the vectors for $\mathit{chicago~state~ university}$ and $\mathit{chicago}$. 

Motivated by this feature of the \ac{CLSM} vectors, query reformulations are represented as the offset vector from the source query to target query \citep{2015_Mitra_ExploringSessionContextUsingDistributedRepresentationsofQueriesandReformulations}. Clustering of query reformulations represented by the offset vectors yields clusters that contain pairs with similar intent transitions. For instance, the query reformulations in which there is an intent jump like \emph{avatar dragons} $\to$ \emph{facebook}, are observed grouped in a cluster. 

\citet{2015_Mitra_ExploringSessionContextUsingDistributedRepresentationsofQueriesandReformulations} uses \ac{CLSM} vectors to define features to represent the session context and rank suggestion candidates against a prefix in personalized query auto-completion. The reformulation feature is the offset vector between the previous query of the user and suggestion candidate. The similarity features are composed of the similarity score of the suggestion candidate to each previous query in the session. For learning to rank the candidate queries against a session, reformulation features and similarity features are included as additional features. The reformulation feature is shown to be provide bigger improvements in the MMR scores than  the similarity features. A combination of the reformulation and similarity features yields the highest MMR scores.

\subsubsection{Learn to generate context}
\label{sec:query-learn-to-generate}

The \ac{HRED} model \cite{2015_Sordoni_AHierarchicalRecurrentEncoder-DecoderforGenerativeContext-AwareQuerySuggestion} is a neural model designed for generating the next query based on the session context vector. The model is able to learn representations for queries, words and sessions simultaneously. The model consists of two encoder RNNs and a decoder RNN. The first encoder RNN maps each query to a distributed representation and the second encoder maps the sequence of query vectors into a session context vector. The decoder generates the most likely query to follow the session, based on the distributed session representation.

The \ac{HRED} is evaluated in two settings. First of all, likelihood scores of the model for a query to follow a session context are used as an additional feature in a learning-to-rank framework to rank query candidates obtained from a co-occurrence based suggestion model. Experimental results show that when the likelihood score is included as a feature, the performance of the baseline ranker is significantly improved. As the second experimental setting, the authors conduct a user study to evaluate the synthetic suggestions generated by the \ac{HRED} model. Users are asked to classify the suggested queries as \emph{useful}, \emph{somewhat useful}, \emph{not useful} categories. A total of 64\% of the queries generated by the \ac{HRED} model was found to be either useful or somewhat useful by users. This score is higher than all the other baselines where the highest score for ``useful or somewhat useful'' is about 45\%.


\section{Ad retrieval}
\label{sec:ads}

We turn to the third and final task considered in this survey: ad retrieval. Here we have two groups of papers: \emph{learn to match} and \emph{learn to predict}.

\subsection{Learn to match}
\label{sec:ads-learn-to-match}

In Table~\ref{tab:ad-learn-to-match}, neural learn to match models for ad retrieval are summarized with the type of \ac{TTU} pairs and the type of the neural network utilized as the \ac{SCN}.

\begin{table}[h!]
\centering
\caption{Neural learn to match models for ad retrieval in web search.}
\label{tab:ad-learn-to-match}
\begin{tabular}{p{2.5cm}p{2.5cm}p{5cm}}
\toprule
\textbf{Study} & \textbf{Pair type} & \textbf{SCN}  \\\midrule
Deep intent~\citep{2016_Zhai_AttentionBasedRecurrentNeuralNetworksforOnlineAdvertising, 2016_Zhai_DeepIntent} & Query-Ad  & Bidirectional RNN/LSTM + Attention Module  \\
\midrule
\cite{2015_Azimi_AdsKeywordRewritingUsingSearchEngineResults} & Ads Keyword Pair  & Utilizes \ac{DSSM} \\
\bottomrule
\end{tabular}
\end{table}

\citet{2015_Azimi_AdsKeywordRewritingUsingSearchEngineResults} use a \ac{DSSM} model for ad keyword re-writing. In paid search, each ad is associated with a set of keywords called \emph{bided keywords}. The ads are ranked against a query and the ads at high rank are displayed to the user. In order to overcome the vocabulary mismatch between user queries and bided keywords, bided keywords are replaced with more common keywords. A set of candidate keywords are extracted from the set of documents returned by a search engine in response to the bided keyword query. The \ac{DSSM} model is leveraged to rank the candidate keywords against the original keywords. 

The \emph{deep-intent} model proposed in~\citet{2016_Zhai_AttentionBasedRecurrentNeuralNetworksforOnlineAdvertising, 2016_Zhai_DeepIntent} comprises a Bidirectional Recurrent Neural Network (BRNN) combined with an attention module as the SCN. The attention module, first introduced in~\cite{2014_NMT_With_Attention} for neural machine translation, is referred to as \emph{attention pooling layer}. This is the first work that employs an attention module for a web search task. A recurrent neural network takes a sequence of words and generates a sequence of distributed representations, so-called \emph{context-vectors}, aligned with each word. Each context vector encodes the semantics of the context from the start to the corresponding word. A Bidirectional Recurrent Neural Network (BRNN) processes the input word sequence in both forward and backward directions. Context vectors generated by a BRNN encode the context after and before the associated word. The pooling strategy is merging the context vectors vectors into a single vector that encodes the semantics of the whole sequence. The sequence vector is assigned the last context vector in \emph{last pooling}, whereas an element-wise max operation is applied on context vectors in \emph{max-pooling}. In \emph{attention pooling}, the sequence vector is computed by a weighted sum of context vectors where the weights are determined by the attention module. The attention module takes a sequence of context vectors and outputs a weight for each vector.
 
The similarity score between a query and an ad is obtained by the dot product of their distributed representations. Similar query-ad pairs for training are extracted from the click logs. Query-ad pairs that have a click are selected as training samples and for each such sample, a randomly selected set of query-ad pairs (without click) are used as negative samples.

Distributed representations are evaluated on click-logs from a product ad search engine which 966K pairs manually labeled by human judges. In the experiments, models that are built from different choices of RNN type (RNN, Bidirectional RNN, LSTM and LSTM-RNN) and pooling strategy (max-pooling, last pooling and attention pooling) are compared. The attention layer provides a significant gain in the AUC (area-under-curve of the receiver operating characteristic) scores when used with RNN and BRNN whereas it performs on a par with last pooling when used with LSTM-based networks. This can be attributed to the power of LSTM units for capturing long-term contextual relations. 

Besides the evaluation of distributed representations for matching, the attention scores are used to extract a subset of query words. The words that have the highest attention scores are selected to rewrite the query. The (query-subquery) pairs are classified by human judges into five categories \textit{same}, \emph{superset}, \emph{subset}, \emph{overlap}, \emph{disjoint}. The sub-queries obtained by the attention module are shown to be of good quality, beating competitive baselines for the majority of queries. 

\subsection{Learn to predict}
\label{sec:ads-learn-to-predict}

In Table \ref{tab:ad-learn-to-predict} learn to predict models for ad retrieval are summarized with the type of \ac{TTU} pairs and the type of the neural network utilized as the \ac{SCN}.

\begin{table}[t]
\centering
\caption{Learn to predict context models for ad retrieval in web search.}
\label{tab:ad-learn-to-predict}
\begin{tabular}{p{2cm}p{1cm}p{3cm}p{4cm}}
\toprule
\textbf{Study} & \textbf{TTU} &  \textbf{Input} &\textbf{Context to predict}  \\
\midrule
context2vec~\cite{2015_Grbovic_Context-andContent-awareEmbeddingsforQueryRewritinginSponsoredSearch} & Query & A query & Other queries (and clicked ads) in the session \\\midrule
context-content2vec~\cite{2015_Grbovic_Context-andContent-awareEmbeddingsforQueryRewritinginSponsoredSearch} & Query & (1) A query (2) The query prefix  & (1) Other queries (and clicked ads) in the session (2) The last word of the query \\ \bottomrule
\end{tabular}
\end{table}

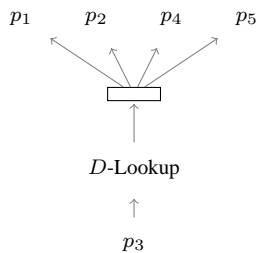
\begin{figure}[t]
  \centering

\begin{tikzpicture}[shorten >=1pt,->,draw=black!50, node distance=\layersep,transform shape,rotate=90]  
    \tikzstyle{every pin edge}=[<-,shorten <=1pt]
    \tikzstyle{neuron}=[circle,draw=black,minimum size=17pt,inner sep=0pt]
    \tikzstyle{vector}=[rectangle,draw=black,minimum width=5pt, minimum height=20pt,inner sep=0pt]
    \tikzstyle{layer}=[rectangle,draw=black,minimum width=5pt, minimum height=50pt,inner sep=0pt]
    \tikzstyle{subnet}=[rectangle,draw=black,minimum size=20pt,inner sep=0pt]
    \tikzstyle{justtex}=[rectangle,minimum size=20pt,inner sep=0pt]
    \tikzstyle{input neuron}=[neuron, fill=green!50];
    \tikzstyle{output neuron}=[neuron, fill=red!50];
    \tikzstyle{hidden neuron}=[neuron, fill=blue!50];
    \tikzstyle{annot} = [text width=4em, text centered]
    \tikzset{hoz/.style={rotate=-90}}   
          
   \path node[justtex,hoz,node distance=2cm] (P-1) at (3,4.5) {$p_1$};
   \path node[justtex,hoz,node distance=2cm] (P-2) at (3,3.5) {$p_2$};
   \path node[justtex,hoz,node distance=2cm] (P-4) at (3,2.5) {$p_4$};
   \path node[justtex,hoz,node distance=2cm] (P-5) at (3,1.5) {$p_5$};

   \path node[justtex,hoz,node distance=2cm] (D) at (0, 3) {$p_3$};
   \path node[justtex,hoz,node distance=2cm] (PEM) at (1, 3) {$D$-Lookup};
   
   \path (D) edge (PEM);
   \path node[vector] (E-P) at (2,3) {};
   \path (PEM) edge (E-P);   
   \path (E-P) edge (P-1);
   \path (E-P) edge (P-2);
   \path (E-P) edge (P-4);
   \path (E-P) edge (P-5);

\end{tikzpicture}
  	\caption{The context2vec model, after~\citep{2015_Grbovic_Context-andContent-awareEmbeddingsforQueryRewritinginSponsoredSearch}.}
  	\label{fig:context2vec}
\end{figure}

The \emph{context-content2vec} model~\cite{2015_Grbovic_Context-andContent-awareEmbeddingsforQueryRewritinginSponsoredSearch}, which has the same architecture as the HDV model depicted in Figure \ref{fig:hdv}, is aimed at learning query embeddings. The context of the query is defined both by its content and other queries in the same session. For the \emph{context-content2vec} model, $p_1,p_2,p_4,p_5$ stands for queries that occur in the same session with $p_3$. Besides the \emph{context-content2vec} model, the \emph{context2vec} model illustrated in Figure \ref{fig:context2vec}, which predicts only temporal context, is also introduced in \cite{2015_Grbovic_Context-andContent-awareEmbeddingsforQueryRewritinginSponsoredSearch}.

The models are trained on Yahoo search logs that contain 12 billion sessions and embeddings for approximately 45 million queries are learned. Learned query embeddings are leveraged for rewriting queries in order to improve search re-targeting. The original query is expanded with its $k$ nearest neighbor queries in the query embedding space. The learned model is evaluated on TREC Web Track 2009--2013 queries and an in-house data set from Yahoo. For queries in the TREC data set, the query rewrites obtained by the proposed models are editorially judged. The \ac{PV-DM} model that only predicts context yields lower editorial grades than the Query Flow Graph (QFG) baseline. Rewrites by \emph{context2vec} and \emph{context-content2vec} embeddings outperform the baseline. The rewrites by the \emph{context-content2vec}$_{ad}$ model, which extends \emph{context-content2vec} by adding the ads and links clicked in the same session to the \ac{TTU} context, are assigned the highest editorial grades on average.


\section{Conclusion}

The purpose of this survey is to offer an introduction to neural models for semantic matching in web search. To this end we reviewed and classified existing work in the area. We used a taxonomy in which we recognize different \acfp{TTU}, different types of usage of learned text representations (``\emph{usage}''), as well as different methods for building representations (``\emph{how}''). Within the latter identified two sub-categories: the \emph{aggregate} and \emph{learn} categories. The \emph{aggregate} category includes methods based on pre-computed word embeddings for computing semantic similarity, while the \emph{learn} category covers the neural semantic compositionality models.

Within the \emph{aggregate} category we observed two major patterns of exploiting word embeddings. In the \emph{explicit} type of use of embeddings, \acp{TTU} are associated with a representation in the word embedding space and semantic similarity of \acp{TTU} is computed based on these representations. In the \emph{implicit} type of use, similarity of word embeddings is plugged in as term similarity in an existing statistical language modeling frameworks for retrieval. Several strategies for adapting word embeddings for document retrieval have been introduced, such as topically constraining the document collection, new similarity functions and the inclusion of TF-IDF weights for aggregating word embeddings. This may be understood as an indication that we need to design \ac{IR} specific objectives for learning distributed representations. Are the training objective and semantic relationships encoded by the embedding vectors useful for the target retrieval task? A future direction would be to identify the types of semantic word relations that matter to semantic matching in web search, across multiple tasks.

We classified the neural semantic compositionality models reviewed in the \emph{learn} category, into the three sub-categories \emph{neural learn to match}, \emph{learn to predict context} and \emph{learn to generate context}, considering the training objectives optimized for learning representations. \emph{Neural learn to match} models are trained using noisy relevance signals from click information in click-through logs whereas the models in the other categories are designed to predict or generate task-specific context of TTUs. Majority of the learn to match and learn to predict models are evaluated on data sets extracted from commercial search engine logs. A comparative evaluation of models from different sub-categories, on publicly available data sets, is required in order to gain a deeper understanding of semantic compositionality for matching. 

Currently, existing learn to predict/generate context models, mostly rely on temporal context windows. In general, it would be interesting to examine other types of contextual relations in search logs, such as long term search history of users and noisy relevance signals exploited by learn to match models. Another future direction would be applications of the attention mechanism~\cite{2014_NMT_With_Attention} for designing models that can predict where a user would attend in document, given a query. 

Looking forward, we believe there are several key directions where progress is needed. First, we presented the document retrieval, query suggestion and ad retrieval tasks as largely disjoint tasks. However, the models proposed for one task may be useful for another. For instance, the \emph{context-content2vec} model of \cite{2015_Grbovic_Context-andContent-awareEmbeddingsforQueryRewritinginSponsoredSearch} was evaluated only on matching ads to queries yet the distributed query representations could also be evaluated for query suggestion or query auto completion~\cite{cai-survey-2016}. In particular, there is a need to compare distributed query representations and similarity/likelihood scores produced by the proposed models on query tasks. In some work, the representations were used as features in learning to rank frameworks and there are no clues about the power of these representations in capturing semantics.

More broadly, there is a need for systematic and broad task-based experimental surveys that focus on comparative evaluations of models from different categories, but for the same tasks and under the same experimental conditions, very much like the reliable information access (RIA) workshop that was run in the early 2000s to gain a deeper understanding of query expansion and pseudo relevance feedback \cite{harman-reliable-2009}.

Another observation is that recently introduced generative models---mostly based on recurrent neural networks---can generate unseen (synthetic) textual units. The generated textual units have been evaluated through user studies \cite{2015_Sordoni_AHierarchicalRecurrentEncoder-DecoderforGenerativeContext-AwareQuerySuggestion,2016_Lioma_Deep_Learning_Relevance}. For the query suggestion task, generated queries have been found to be useful; and so have word clouds of a synthetic document. The impact of these recent neural models on user satisfaction or retrieval scenarios should be investigated on real scenarios.

Finally, over the years, web search has made tremendous progress by learning from user behavior, either by introducing, e.g., click-based rankers~\cite{radlinski-query-2005} or, more abstractly, by using models that capture behavioral notions such as examination probability and attractiveness of search results through click models~\cite{chuklin-click-2015}. How can such implicit signals be used to train neural models for semantic matching in web search? So far, we have only seen limited examples of the use of click models in training neural models for web search tasks.


\appendix

\section{Acronyms used}

\begin{acronym}[MMMMMM]
\acro{AI}{artificial intelligence}
\acro{BoEW}{Bag of Embedded Words}
\acro{BWESG}{Bilingual word Embeddings Skip-Gram}
\acro{CBOW}{Continuous Bag of Words}
\acro{CLSM}{Convolutional Latent Semantic Model}
\acro{DESM}{Dual Embedding Space Model}
\acro{DFR}{Divergence From Randomness}
\acro{DSM}{distributional semantic model}
\acro{DSSM}{Deep Structured Semantic Model}
\acro{EPV}{Extended Paragraph Vector}
\acro{GLM}{Generalized Language Model}
\acro{HAL}{Hyperspace Analog to Language}
\acro{HDV}{Hierarchical Document Vector}
\acro{HRED}{Hieararchical Recurrent Encoder Decoder}
\acro{IR}{information retrieval}
\acro{IS}{Importance Sampling}
\acro{LDA}{Latent Dirichlet Allocation}
\acro{LM}{language model}
\acro{LSA}{Latent Semantic Analysis}
\acro{LSI}{Latent Semantic Indexing}
\acro{NCE}{Noise Contrastive Estimation}
\acro{NEG}{Negative Sampling}
\acro{NLM}{Neural Language Model}
\acro{NLTM}{Neural Translation Language Model}
\acro{NNLM}{Neural Network Language Model}
\acro{PLSA}{Probabilistic Latent Semantic Analysis}
\acro{PV}{Paragraph Vector}
\acro{PV-DBOW}{Paragraph Vector with Distributed Bag of Words}
\acro{PV-DM}{Paragraph Vector with Distributed Memory}
\acro{QLM}{query language model}
\acro{RNNLM}{Recurrent Neural Network Language Model}
\acro{SC}{semantic compositionality}
\acro{SCN}{Semantic Compositionality Network}
\acro{SGNS}{Skip-Gram with Negative Sampling}
\acro{TTU}{target textual unit}
\acro{WMD}{Word Mover's Distance}
\end{acronym}

\bibliographystyle{spbasic}
\bibliography{refs}

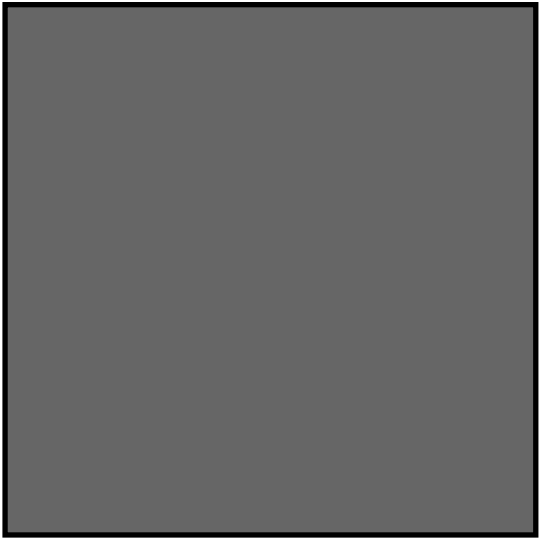
\end{document}